\documentclass[bibyear]{aa}
\usepackage{graphicx}
\usepackage{amsmath}
\usepackage{txfonts}
\usepackage[]{natbib}
\begin{document} 

     \title{The first frost in the Pipe Nebula\thanks{Based on
         data collected by SpeX at the Infrared Telescope
         Facility, which is operated by the University of Hawaii
         under contract NNH14CK55B with the National Aeronautics
         and Space Administration.}\fnmsep \thanks{Based also on
         data obtained at the W.M. Keck Observatory, which is
         operated as a scientific partnership among the
         California Institute of Technology, the University of
         California, and the National Aeronautics and Space
         Administration. The Observatory was made possible by
         the generous financial support of the W.M. Keck
         Foundation.}}


   \author{Miwa Goto\inst{1,2}, 
           J. D. Bailey \inst{1},
           Seyit Hocuk\inst{1},
           Paola Caselli\inst{1},
           Gisela B. Esplugues\inst{1,3},
           Stephanie Cazaux\inst{3,4,5},
           Marco Spaans\inst{3}}

   \institute{Max-Planck-Institut f\"ur extraterrestrische
              Physik, Giessenbachstrasse 1, 85748, Germany 
   \and
              Universit\"ats-Sternwarte M\"unchen,
              Ludwig-Maximilians-Universit\"at, 
              Scheinerstrasse 1, 81679 M\"unchen, Germany \\
              \email{mgoto@usm.lmu.de}
   \and
              Kapteyn Astronomical Institute, University of
              Groningen, PO Box 800, 9700 AV Groningen, The
              Netherlands
   \and
              Leiden Observatory, Leiden University, P.O.
              Box 9513, 2300 RA Leiden, The Netherlands
   \and 
              Aerospace Engineering, Delft University of Technology, 
              Kluyverweg 1, 2629 HS Delft, The Netherlands
   }

   \date{\today}

  \abstract 
      {Spectroscopic studies of ices in nearby star-forming
        regions indicate that ice mantles form on dust grains in
        two distinct steps, starting with polar ice formation
        (H$_2$O rich) and switching to apolar ice (CO rich).}
  {We test how well the picture applies to more diffuse and
    quiescent clouds where the formation of the first layers of
    ice mantles can be witnessed.}
  {Medium-resolution near-infrared spectra are obtained
        toward background field stars behind the Pipe Nebula.}
  {The water ice absorption is positively detected at
    3.0\,$\mu$m in seven lines of sight out of 21 sources for which
    observed spectra are successfully reduced. The peak optical
    depth of the water ice is significantly lower than those in
    Taurus with the same $A_V$. The source with the highest
    water-ice optical depth shows CO ice absorption at
    4.7\,$\mu$m as well. The fractional abundance of CO ice with
    respect to  water ice is 16$^{+7}_{-6}$\,\%, and about
    half as much as the values typically seen in low-mass
    star-forming regions.}
{A small fractional abundance of CO ice is consistent with some of
  the existing simulations.
  Observations of CO$_2$ ice in the early diffuse phase of a
  cloud play a decisive role in understanding the switching
  mechanism between  polar and  apolar ice formation.}



\titlerunning{Water and CO Ice in the Pipe Nebula}
\authorrunning{Goto et al.}

\maketitle
   \begin{figure*}
     \includegraphics[width=\textwidth]{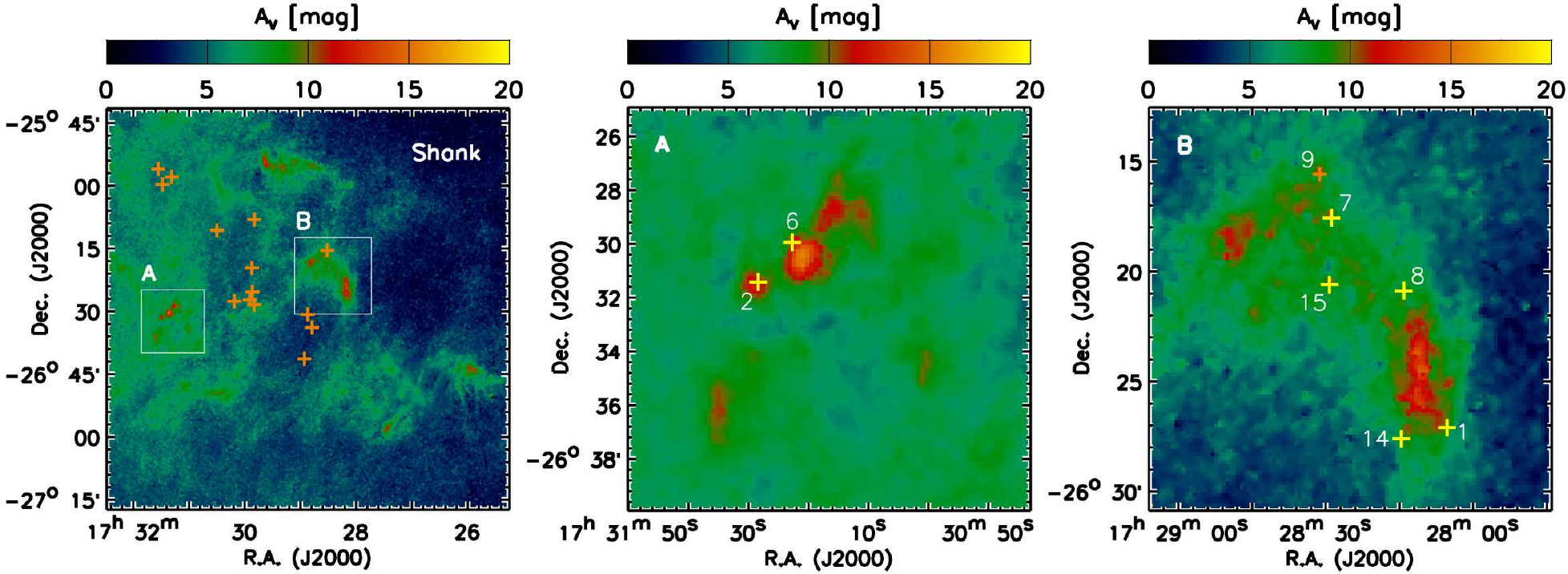}
     \caption{Extinction map of the Shank in the Pipe Nebula
       based on the near-infrared color excess
       \citep{Roman-Zuniga:2010ApJ...725.2232R}. Overlaid are
       the positions of the targets. The sources where water ice
       is positively detected are marked by yellow crosses,
       while those with negative detections are marked by orange
       crosses. The zoom-in view of the positive detections at
       the regions A and B are shown in the middle and  right
       panel, respectively.} \label{p5}
   \end{figure*}
\section{Introduction}

 Star formation starts with a diffuse interstellar cloud that
 gradually condenses into a dark molecular cloud. Dense cores
 may form inside the dark cloud and a fraction of them
 eventually collapse into new stellar systems. The piled-up
 material locally increases the visual extinction, then
 photodissociation rates drop exponentially, and molecules start
 to thrive  \citep[e.g.,][]{Tielens:2005pcim.book.....T}. Dust
 grains in the dark clouds are generally cold \citep[$<15$\,K;
 ][]{Spitzer:1998ppim.book.....S,Hocuk:2017arXiv170402763H} if
 the clouds are devoid of active star formation.  As soon as the
 visual extinction reaches $A_V=2$--$3$\,mag, water molecules
 copiously form on the grain surfaces via successive
 hydrogenation of atomic oxygen
 \citep[e.g.,][]{Aikawa:2003ApJ...593..906A,Hollenbach:2009ApJ...690.1497H}. Other
 molecules in the gas phase, in particular CO, stick onto dust
 surfaces, producing mixed and layered ice mantles on the
 grains \citep[][]{Chiar:1995ApJ...455..234C}. These mantles
 are the hideouts of the missing molecules in the gas
 \citep[e.g.,][]{Caselli:1999ApJ...523L.165C}, incubators of
 complex organic molecules
 \citep[e.g.,][]{Vasyunin:2013ApJ...762...86V,Vasyunin:2017ApJ...842...33V},
 and adhesives that facilitate  grain growth up to the size of 
 planetesimals \citep[e.g.,][]{Gundlach:2015ApJ...798...34G}.

The {\em Spitzer} Space Telescope brought a leap forward in the
study of the ices in the interstellar medium \citep[see review in][]{Boogert:2015.53.}. The superb sensitivity of the
spectrograph enabled observations of faint field stars behind
star-forming regions as background continuum sources
\citep[e.g.,][]{Knez:2005ApJ...635L.145K,Chiar:2011ApJ...731....9C,Boogert:2013ApJ...777...73B}.
The wide wavelength coverage facilitated the comparative study
of the molecules in the ice. With timely advances in laboratory
experiments
\citep[e.g.,][]{Ioppolo:2008ApJ...686.1474I,Oba:2010ApJ...712L.174O,
  Noble:2011ApJ...735..121N} and numerical simulations
\citep[e.g.,][]{Aikawa:2003ApJ...593..906A,Taquet:2012A&A...538A..42T,Cuppen:2013ChRv..113.8840C,Sipila:2013A&A...554A..92S,Hocuk:2016MNRAS.456.2586H},
our understanding of the formation of ices has been
substantially refined. A picture that gradually emerged is the
formation of the ice mantle taking place in two sequential but
distinct stages
\citep[e.g.,][]{Oberg:2011ApJ...740..109O,Boogert:2015.53.}.
The first stage is characterized by a rapid formation of water
ice along with CO$_2$, which  may form either by the surface
reaction of CO and OH \citep{Ruffle:2001MNRAS.324.1054R} or by
cosmic-ray bombardment of water ice on top of carbonaceous
material \citep{Mennella:2004ApJ...615.1073M}. The second stage
is the accumulation of CO ice. The catastrophic freeze-out of CO
is often observed toward dense cores at the volume density above
a few $\times$10$^4$\,cm$^{-3}$
\citep{Caselli:1999ApJ...523L.165C,Tafalla:2002ApJ...569..815T,Tafalla:2006A&A...455..577T}.

The motivation of the present observations is to collect
empirical insights into the first steps of  ice formation in
a quiescent molecular cloud away from the sites of active star
formation without complications of (proto-)stellar feedback. Our
target is the Pipe Nebula. Compared to other nearby star-forming
regions such as Taurus, Lupus, and Serpens, the Pipe Nebula is
clearly more quiescent. Only a handful of embedded sources are
known in the entire complex, most of which are located in B\,59
at the tip of the Stem
\citep{Duarte-Cabral:2012A&A...543A.140D,Hara:2013ApJ...771..128H}.
The sub-mm molecular lines are narrow
\citep[0.2--0.5\,km\,s$^{-1}$, ][]{Muench:2007ApJ...671.1820M,
  Frau:2010ApJ...723.1665F}, indicating that the gas is mostly
subsonic \citep{Lada:2008ApJ...672..410L} and giving support to
the early evolutionary stage of the cloud. The Pipe Nebula is
located at 130--145\,pc  from the Sun
\citep{Lombardi:2006A&A...454..781L,Alves:2007A&A...470..597A}.
Projection against the Galactic Bulge guarantees a large
reservoir of background stars to choose from.

We illustrate in \S\,\ref{obsdata} how the spectroscopic data
are collected, and the optical depths and the column densities
of water and CO ices are extracted. The caveats in using $A_V$
as a reference of the optical depths of ice is discussed in
\S\,\ref{A_V}. Possible interpretations of the observations with
locally enhanced radiation field and chemical evolution of the
ice are discussed in \S\,\ref{rad} and \S\,\ref{chemevo},
respectively. The column densities of water, CO, and CO$_2$ ice
are complemented by the literature, and are compared to the latest
simulations in \S\,\ref{simcom}. The summary of the present
study is given in \S\,\ref{summary}.

\section{Observations, data reductions, and results \label{obsdata}}
\subsection{SpeX/IRTF}
\subsubsection{Observation}

The near-infrared spectrograph SpeX
\citep{Rayner:2003PASP..115..362R} at the IRTF on Mauna Kea is
used to collect the medium resolution spectra of the field stars
behind the Pipe Nebula. SpeX is equipped with a prism
cross-disperser. Together with the large-format
2048\,$\times$\,2048 pixel detector array, the instrument covers
a wide wavelength interval with a single exposure. The wide
simultaneous wavelength coverage is an integral component of the
present observations; the biggest challenge in observing water
ice from the ground is to constrain the short wavelength
continuum properly. The stretching vibrational band of water ice
is centered at 3.0\,$\mu$m and the absorption profile is
broad. The absorption extends beyond 2.8\,$\mu$m in the short
wavelength, overlapping with the atmospheric water vapor at
2.7\,$\mu$m. It is therefore hard to set the continuum level
correctly with the $L$-band spectroscopy (2.8--4.1\,$\mu$m)
only. Spectroscopy or photometry at the $K$-band
(1.9--2.5\,$\mu$m) can be used to set the continuum baseline,
which is not only time consuming in observations, but also
introduces other systematics such as variable slit
throughput. Simultaneous coverage of $K$ and $L$ by SpeX is a
critical advantage to the observational study of water ice from
the ground.

All targets are from the Shank in the Pipe Nebula, where 
 no signs of star formation have been identified to date.  The stars
were selected so that they collectively cover the visual
extinction from $A_V$=3 to 12\,mag (Fig.~\ref{p5}), referring to
the extinction map of \cite{Roman-Zuniga:2010ApJ...725.2232R}.
The selected targets are brighter than $K_s<$9\,mag in 2MASS
\citep{Skrutskie:2006AJ....131.1163S} and $W1<8.6$\,mag
(3.4\,$\mu$m) in WISE photometry
\citep{Wright:2010AJ....140.1868W}. Visually bright stars in
the USNO B1.0 catalog ($R<13$\,mag) are filtered out, because
the chances are high that they are foreground stars. The summary
of the targets is given in Table~\ref{t1}.

\begin{table*}
\begin{center}
\scriptsize

\caption{Summary of targets and optical depths of ices\label{t1}}
\begin{tabular}{rl c r cc cc cc cc}
\hline \hline
\multicolumn{1}{c}{ID \#}&\multicolumn{1}{c}{2MASS ID}&\multicolumn{1}{c}{$K_s$}&\multicolumn{1}{c}{Spec. Type}&\multicolumn{1}{c}{$A_V$ (CE)}&\multicolumn{1}{c}{$A_K$ (CE)}&\multicolumn{1}{c}{$A_K$ (SpeX)}&\multicolumn{1}{c}{$A_K$ (SED)}&\multicolumn{1}{c}{$\tau_{3.0}$}&\multicolumn{1}{c}{$N$(H$_2$O)}&\multicolumn{1}{c}{$\tau_{4.7}$}&\multicolumn{1}{c}{$N$(CO)\tablefootmark{a}}\\
\multicolumn{1}{c}{}&\multicolumn{1}{c}{}&\multicolumn{1}{c}{[mag]}&\multicolumn{1}{c}{(SpeX)}&\multicolumn{1}{c}{[mag]}&\multicolumn{1}{c}{[mag]}&\multicolumn{1}{c}{[mag]}&\multicolumn{1}{c}{[mag]}&\multicolumn{1}{c}{}&\multicolumn{1}{c}{[10$^{17}$\,cm$^{-2}$]}&\multicolumn{1}{c}{}&\multicolumn{1}{c}{[10$^{16}$\,cm$^{-2}$]}\\
\hline
 2 & 17312818-2631268 &  8.913 &         M5.5III & 11.97 &  1.34 &  1.32 &  1.09 &  0.36$^{+0.08}_{-0.02}$ &  6.05$^{+1.27}_{-0.32}$ & 0.13$\pm$0.05 & 9.7$\pm$3.7\tablefootmark{b} \\
 6 & 17312249-2629585 &  8.256 &           M6III & 11.21 &  1.26 &  2.63 & --- &  0.29$^{+0.09}_{-0.03}$ &  4.89$^{+1.48}_{-0.52}$ & $<$0.11 & $<$8.9 \\
15 & 17282929-2620358 &  8.404 &         M5.5III & 10.04 &  1.12 &  1.13 &  0.99 &  0.23$^{+0.08}_{-0.02}$ &  3.93$^{+1.27}_{-0.38}$ & $<$0.12 & $<$9.0 \\
 1 & 17280535-2627053 &  8.982 &           M3III & \phantom{1}9.54  &  1.07 &  1.13 &  0.96 &  0.25$^{+0.02}_{-0.02}$ &  4.26$^{+0.31}_{-0.39}$ & $<$0.11 & $<$8.3 \\
 9 & 17283130-2615350 &  8.156 &           M3III & \phantom{1}9.52  &  1.07 &  0.99 & --- & $<$0.24 & $<$3.97 & $<$0.62 & $<$33.8 \\
 8 & 17281419-2620522 &  8.491 &         M5.5III & \phantom{1}9.52  &  1.07 &  1.21 &  1.32 &  0.18$^{+0.08}_{-0.02}$ &  2.96$^{+1.31}_{-0.39}$ & $<$0.11 & $<$8.4 \\
 7 & 17282887-2617337 &  8.504 &           M6III & \phantom{1}9.10  &  1.02 &  1.85 &  1.03 &  0.23$^{+0.05}_{-0.09}$ &  3.88$^{+0.91}_{-1.48}$ & $<$0.09 & $<$7.0 \\
14 & 17281466-2627361 &  7.844 &         M5.5III & \phantom{1}8.85  &  0.99 &  1.10 &  1.13 &  0.18$^{+0.04}_{-0.02}$ &  3.07$^{+0.68}_{-0.31}$ & $<$0.16 & $<$12.5 \\
24 & 17294935-2628335 &  7.468 &         M5Ib-II & \phantom{1}6.22  &  0.70 &  1.04 &  0.69 & $<$0.23 & $<$3.79 & $<$0.19 & $<$10.6 \\
31 & 17311682-2557571 &  7.896 &         M5Ib-II & \phantom{1}6.02  &  0.67 &  1.03 &  0.89 & $<$0.63 & $<$10.61 & $<$0.25 & $<$13.8 \\
34 & 17313129-2556041 &  7.286 &         M5Ib-II & \phantom{1}6.01  &  0.67 &  0.75 &  0.79 & $<$0.23 & $<$3.89 & $<$0.30 & $<$16.6 \\
50 & 17294916-2608152 &  7.654 &         M5Ib-II & \phantom{1}6.01  &  0.67 &  0.76 &  1.47 & $<$0.18 & $<$3.09 & $<$0.28 & $<$15.2 \\
39 & 17302893-2610478 &  7.961 &         M5Ib-II & \phantom{1}5.74  &  0.64 &  0.85 &  0.80 & $<$0.60 & $<$10.04 & $<$0.33 & $<$18.2 \\
32 & 17312691-2559455 &  7.991 &           M6III & \phantom{1}5.72  &  0.64 &  1.56 & --- & $<$0.43 & $<$7.26 & $<$0.19 & $<$10.6 \\
52 & 17295176-2619482 &  7.841 &         M5.5III & \phantom{1}5.60  &  0.63 &  0.90 &  1.02 & $<$0.27 & $<$4.60 & $<$0.24 & $<$13.3 \\
44 & 17301059-2627445 &  6.795 &         M5Ib-II & \phantom{1}5.31  &  0.59 &  0.89 &  0.76 & $<$0.56 & $<$9.45 & $<$0.21 & $<$11.8 \\
21 & 17295126-2625320 &  7.991 &         M5.5III & \phantom{1}5.14  &  0.58 &  0.70 &  0.67 & $<$0.24 & $<$4.07 & $<$0.28 & $<$15.1 \\
22 & 17295408-2627226 &  7.920 &           M6III & \phantom{1}5.07  &  0.57 &  1.57 & --- & $<$0.47 & $<$7.80 & $<$0.18 & $<$9.9 \\
64 & 17285573-2641305 &  7.644 &         M3.5III & \phantom{1}4.95  &  0.55 &  0.51 &  0.79 & $<$0.34 & $<$5.69 & $<$0.74 & $<$40.3 \\
61 & 17284743-2634008 &  7.792 &           M6III & \phantom{1}3.66  &  0.41 &  1.44 &  0.85 & $<$0.54 & $<$8.97 & $<$0.20 & $<$11.0 \\
57 & 17285211-2630536 &  6.877 &         M5.5III & \phantom{1}3.40  &  0.38 &  0.71 &  0.63 & $<$0.20 & $<$3.37 & $<$0.26 & $<$14.4 \\
\hline
\hline
\end{tabular}
\tablefoot{Ranges given as uncertainty are 1 $\sigma$.
Upper limits are 3 $\sigma$.\\
ID : Numbering of sources given in Fig.~\ref{p5}. The sources are ordered according to  $A_V$ (CE).\\
$A_V$ (CE) : $A_V$ by Rom\'an-Z\'u\~niga et al. (2010) with color excess technique. 
$A_K$ (CE) is calculated by multiplying $A_V$ by the factor 0.112 (Rieke \& Lebofsky 1985).\\
$A_K$ (SpeX) : obtained by comparing the observed SpeX spectra with the templates spectra of IRTF Spectral Library (Rayner et al. 2009).\\
$A_K$ (SED) : obtained  by comparing SED with stellar model by Bressan et al. (2012).\\
$\tau_{\rm 3.0}$ : peak optical depth of water ice at 3.0\,$\mu$m.\\
$\tau_{\rm 4.7}$ : peak optical depth of CO ice at 4.672\,$\mu$m.\\
\tablefoottext{a}{The column density of CO ice.
The CO ice spectra at 4.7\,$\mu$m are obtained by NIRSPEC at Keck II telescope 
for the sources in which the water ice is positively detected (\#2, 6, 15, 1, 8, 7, 14).
Other upper limits are based on SpeX/IRTF observations.}\\
\tablefoottext{b}{Broad absorption solution without the short-wavelength shoulder 4.655--4.665\,$\mu$m included in the continuum (see \S\,\ref{t47}).
If the shoulder is counted as a continuum, the column density is 
reduced to $N$(CO)=$(6.4\pm 3.3)\times 10^{16}$\,cm$^{-2}$.}
}
\end{center}
\end{table*}
\normalsize


The observations were carried out on eight half-nights in May and
June in 2015. The instrument was remotely operated from Munich
in Germany. All spectra were obtained with the optics setting
{\tt LXD\_long} with the 0\farcs8 slit. The slit was roughly
aligned to the parallactic angle to reduce the loss of flux at
the slit by the atmospheric refraction. The spectra were
recorded by nodding the telescope along the slit every second
exposure to remove the sky background emission. Early-type
standard stars were observed immediately after the science
targets. Spectroscopic flat field, i.e., a continuum illumination
of a halogen lamp, was obtained with the same instrument setting
after the observation of the standard stars. Argon lamp spectra
were obtained after the flat field to map the wavelength on the
detector array.

\subsubsection{Extraction of spectra}

The one-dimensional spectra were obtained using {\em xSpexTool}
v4.1, which  runs on an intuitive graphical user interface
\citep{Cushing:2004PASP..116..362C}. The IDL-based spectral
reduction package performs flux calibration, non-linearity
correction, wavelength calibration, coadding of the
two-dimensional spectrograms, aperture extraction, and merging
of the spectra extracted from the different diffraction
orders. More detail is found on the IRTF SpeX webpage\footnote
{http://irtfweb.ifa.hawaii.edu/\textasciitilde spex/}.

The atmospheric absorption features are removed by dividing the
object spectrum by that of a spectroscopic standard star. The
blackbody spectra of the effective temperature of the standard
star is multiplied to restore the continuum slope. The effective
temperatures are taken from Allen's Astrophysical Quantities
\citep{Cox:2000asqu.book.....C} of the corresponding spectral
type documented in {\em SIMBAD}.  Prominent \ion{H}{I}
absorption lines (Br\,$\gamma$, Pf\,$\gamma$, and Br\,$\alpha$)
are removed beforehand by fitting a Lorentzian function and
subtracting it from the spectra. Less prominent \ion{H}{I} lines
blended with the atmospheric absorption lines (e.g., Pf\,$\delta$
at 3.297~$\mu$m) are removed after the object spectra are
divided by the standard star spectra.

The absolute flux calibration is performed by scaling the object
spectra to the broadband photometry at 2MASS $K_s$ and {\em
  WISE} $W1$. The SpeX spectra are convolved by the 2MASS/WISE
filter transmission curves before comparing to the photometry. A
mean scaling factor at $K_s$ and $W1$ bands is multiplied to the
object spectrum without adjusting $K$- and $L$-band spectra to
each other.

   \begin{figure*}
    \center 
   \includegraphics[width=0.77\textwidth]{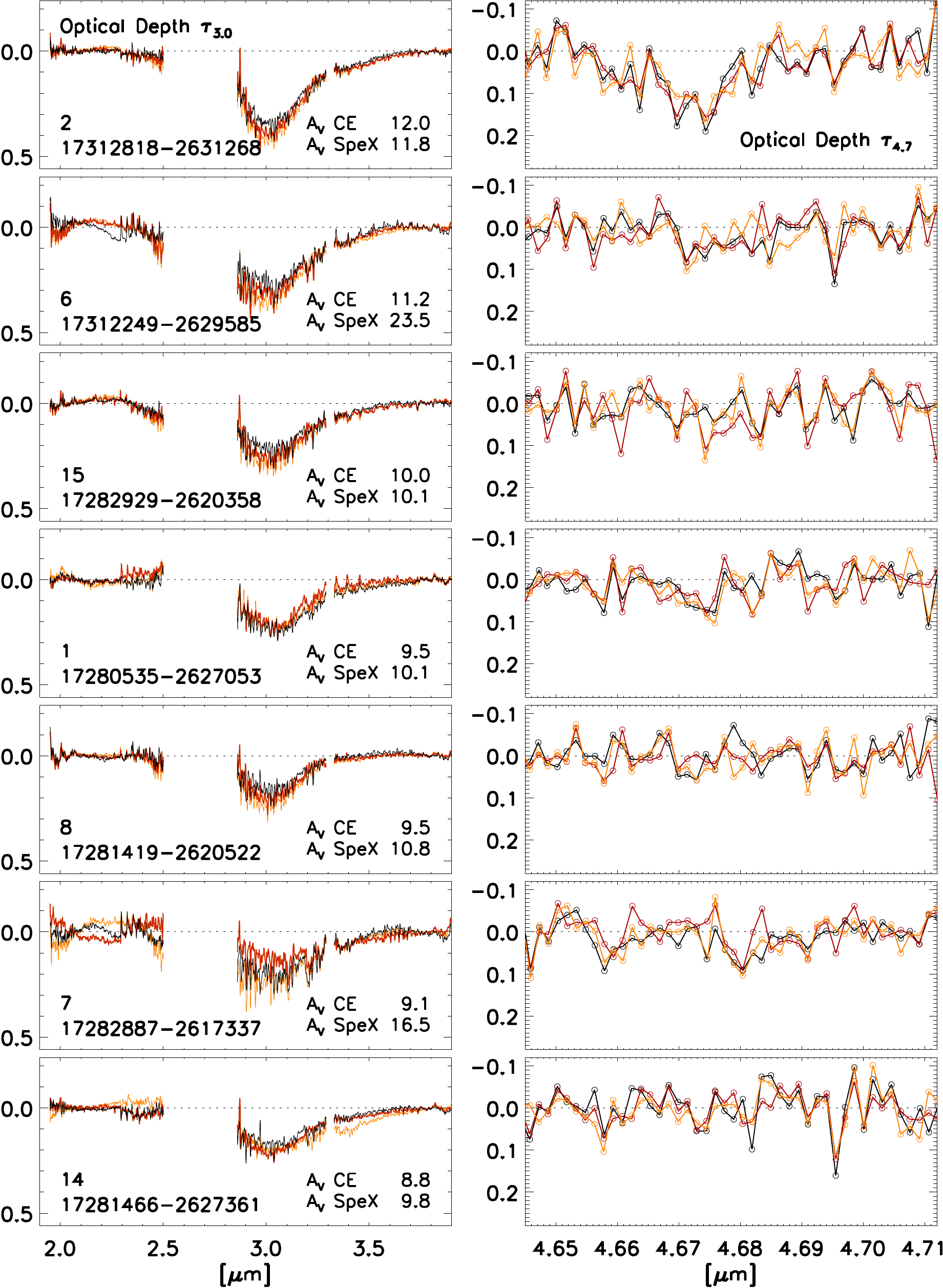}
   \caption{Optical depth spectra of water (3.0\,$\mu$m; left
     column) and CO ice (4.672\,$\mu$m; right column) toward the
     seven sources where water ice was positively detected in
     the observations by SpeX spectrograph at the IRTF. The
     numbering of the sources in Fig.~\ref{p5} and
     Table~\ref{t1} is shown at the bottom left of the water-ice
     panels with the 2MASS IDs. The CO ice spectra were obtained
     by NIRSPEC spectrograph at Keck\,II. The whole SpeX spectra
     including negative detections are compiled in Fig.~\ref{s9}
     in Appendix \ref{ap1}. Black, red, and orange plots, are
     the spectra in which photospheric features are removed by
     referring to the best, the second best, and the third best
     matching $M$-type template stars. The visual extinctions
     based on the color excess \citep[$A_V$~CE;
     ][]{Roman-Zuniga:2010ApJ...725.2232R} and on the template
     matching with the IRTF Spectral Library \citep[$A_V$ SpeX;
     ][]{Rayner:2009ApJS..185..289R} are shown at the bottom
     right of the water ice spectra.
     \label{t1_coi}}
   \end{figure*}
\subsubsection{Template matching}

The targets we observed are late-type  stars with clear
photospheric CO bandhead absorption at the 2.3\,$\mu$m
(Fig.~\ref{s9} in Appendix \ref{ap1}). Following
\cite{Boogert:2011ApJ...729...92B}, the photospheric lines are
removed by comparing the object spectra with those of the
spectroscopic template stars in the IRTF Spectral Library
\citep{Rayner:2009ApJS..185..289R}. A series of template spectra
are reddened by a trial $A_K$ ranging from 0.0 to 4.0\,mag with
a 0.01\,mag step that roughly corresponds to sampling $A_V$ from
0 to 36\,mag by 0.1\,mag step. The empirical infrared extinction
curve obtained by \cite{Boogert:2011ApJ...729...92B} is
used. The reddened template spectrum is scaled to the continuum
of the object spectrum at 3.6--4.0~$\mu$m, and the residual
spectrum against the object spectrum is calculated. The
combination of the template and $A_K$ that minimizes the
residual spectrum is taken as a match. The template spectra
themselves are slightly affected by the extinction on their own.
\cite{Rayner:2009ApJS..185..289R} corrected the extinction in
the library spectra in the case that substantial reddening is
inferred in their $B-V$ colors. The extinction corrected library
spectra are used whenever available. Most of the spectra in the
IRTF Spectral Library are obtained with the 0\farcs3 slit, and
therefore at higher spectral resolution than our
observations. The library spectra are convolved before comparing
to the object spectra. A small wavelength shift between the
template and the observed spectra is corrected down to the order
of 10$^{-4}$~$\mu$m.

The residual spectra are evaluated between 1.9 and 4.1\,$\mu$m
excluding the wavelength interval of the water ice band and the
long-wavelength absorption excess (2.5--3.6\,$\mu$m). The best
matching template spectra are shown with the object spectra in
the middle column of Fig.~\ref{s9}. Out of the 46 targets
observed, 21 stars match the spectral type M6 or earlier, and
are used in further analysis (Table~\ref{t1}). The rest of the
sample consists of stars later than M6, whose photospheric
absorption of water vapor is too deep to be corrected by the
template spectra (not shown in Fig.~\ref{s9}).

\subsubsection{Optical depth and column density}

The optical depth is calculated as the natural logarithm of the
scaled template spectra ($f_0$) divided by the object spectra
($f$), i.e., $\tau_\lambda = -\ln\frac{f}{f_0}$. The optical
depth spectra are fit by a Lorentzian function to measure the
peak absorption depth at 3.0~$\mu$m ($\tau_{\rm 3.0}$). The
choice of the template spectrum is a significant source of
uncertainty in the optical depth. We have selected the three
best matching templates and show the range of the optical depth
spectra in the right panels of Fig.~\ref{s9} to highlight the
systematic uncertainty that stems from the templates. In order
to take into account this uncertainty, the upper and the lower
margins of $\tau_{\rm 3.0}$ are set in such a way that the whole
range of the peak optical depths calculated with the three
different templates are encompassed. The standard deviation of
the continuum baseline is added upon that range by a squared sum
to represent the statistical part of the uncertainty. The seven
sources ( 2, 6, 15, 1, 8, 7, 14) where the peak optical depths
$\tau_{3.0}$ are observed with a significance larger than twice
the total uncertainty are taken as positive detections
(Fig.~\ref{t1_coi}). The selection criterion matches the
positive detections by visual inspection as well.

The equivalent widths of the water ice are calculated by scaling
the ice absorption spectrum recorded in the laboratory
\citep{Hudgins:1993ApJS...86..713H} to the peak optical depth of
the observed spectrum. The excess absorption at the long
wavelength shoulder at 3.4\,$\mu$m is therefore excluded from
the integration of the water ice absorption. The equivalent
width is converted to the unit of wavenumber cm$^{-1}$ and
divided by the integrated band strength
$A=2.0\times10^{-16}$\,cm, adopted from
\cite{Gerakines:1995A&A...296..810G} to calculate the water ice
column density $N({\rm H_2O})$. No CO ice absorption was found
in the SpeX spectra. The results are summarized in
Table~\ref{t1}.


\subsection{NIRSPEC/Keck II}
\subsubsection{Observation}

The seven sources where water ice is positively detected at
3.0\,$\mu$m by SpeX are further investigated using NIRSPEC
spectrograph \citep{McLean:1998SPIE.3354..566M} at the Keck\,II
telescope. One of the advantages of NIRSPEC over other infrared
spectrographs on 8m class telescopes is that the instrument
delivers relatively high spectral resolution with a wide
slit. This is particularly useful when no wavefront reference
source is available to use an adaptive optics system, as in the
present case. We used a slit 0\farcs72 wide and 24\arcsec\, long
to attain the spectral resolution $R$=15,000. Although the CO
ice band is much broader (FWHM$\sim$0.01\,$\mu$m or
$R\approx$1000 required for a Nyquist sampling), observation
with a high spectral resolution is favored because it
facilitates the removal of the photospheric absorption of
late-type stars. The echelle and the cross-dispersing grating
are set to 61.0\degr~ and 36.7\degr~ to place the spectral strip
of the diffraction order 16 on the lower quadrants of the
detector array with the wavelength 4.672\,$\mu$m set at the
center. The {\tt M-WIDE} filter was put in the optical path in
conjunction. The instrumental setting was fine-tuned by the
NIRSPEC Echelle Format Simulator
(EFS)\footnote{https://www2.keck.hawaii.edu/inst/nirspec/EFS.html}
in advance of the observing runs.

   \begin{figure}
    \center 
     \includegraphics[width=0.45\textwidth]{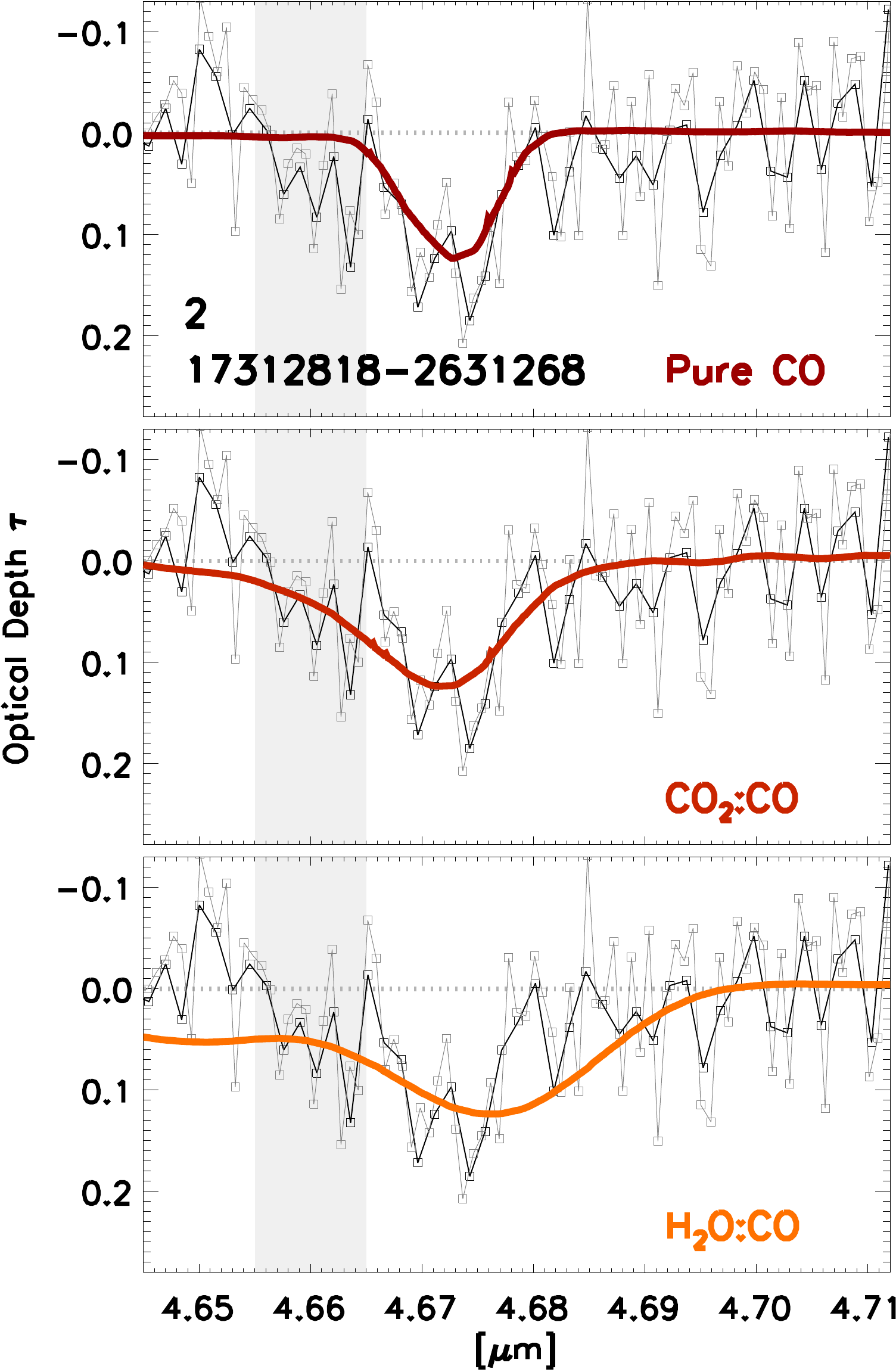}
     \caption{CO ice spectra toward source \#2
       (17312818$-$2631268) recorded by NIRSPEC
       spectrograph. The black plot is the spectra binned by 21
       pixels from the original sampling, and the gray plot by
       11 pixels. The observed spectra shown in the three panels
       are identical. The laboratory spectra of CO ice obtained
       by \citet{Palumbo:1993A&A...269..568P} are overlaid in
       red, dark orange, and light orange in each panel. The top
       panel shows the spectrum of CO ice deposited by itself,
       the middle panel CO ice deposited with CO$_2$, and the
       bottom panel CO ice embedded in water ice. The dotted
       gray lines are the assumed continuum level with zero
       optical depths. The region marked in light gray indicates
       where the continuum level is uncertain. 
       \label{pa}}

   \end{figure}

The observation was carried out in the first half of the nights of 6--8
August 2016 UT. The instrument was operated from Keck
Headquarters in Waimea. The slit was roughly oriented to the
parallactic angle at the acquisition of the target. The spectra
were recorded by moving the source along the slit at every
second exposure. The pointing limit of Keck II is about
37\degr~in elevation on the southern
sky\footnote{https://www2.keck.hawaii.edu/inst/common/TelLimits.html}. The
science targets in the Pipe Nebula were observed in the first
quarter of the nights, and in total 20 $M$-type template stars
were observed in the second quarters. The template spectra were
used to remove the photospheric absorptions of the science
targets. A few early-type standard stars were observed per
night.

\subsubsection{Extraction of spectra}

The preliminary data reduction is performed using {\em REDSPEC}
IDL
package\footnote{https://www2.keck.hawaii.edu/inst/nirspec/redspec.html}. The
raw frames are pair subtracted, and the pixel sensitivities are
normalized by the flat field of the night. The spectral
curvature and the curvature of the slit images are fit by
polynomial functions and rectified. Wavelength dispersion is
mapped out on the detector, referring to the sky emission
lines. One-dimensional spectra are extracted from the
coverture-corrected spectrograms.

The science spectra on the sources behind the Pipe Nebula, and
the template spectra of nearby $M$-type stars are corrected for
the atmospheric absorption lines by dividing by the spectra of
the early-type standard stars observed through similar
airmasses. Small mismatches in the wavelength solution, the
spectral resolution, and the optical depths of the atmospheric
lines are corrected manually. For each science target, the three
best matching $M$-type template stars are identified on the
basis of maximum cross-correlation.  The
  selected template star spectra were used to remove the
  photospheric absorption lines of the science target.

\subsubsection{Optical depth and column density\label{t47}}

Fully reduced spectra are binned by 21 pixels to the spectral
resolution $R$=3600, and shown in Fig.~\ref{t1_coi}. The optical
depth of CO ice is calculated assuming that the wavelength
intervals 4.645--4.655\,$\mu$m and 4.685--4.700\,$\mu$m are
genuine continuum emission with null ice absorption. Looking at
the spectra, CO ice appears to be detected on source \#2
(2MASSJ~17312818$-$2631268). The line of sight to source \#2
passes a dense core with the smallest impact parameters among
the sources observed (Fig.~\ref{p5}), and it shows the highest
water-ice optical depth in the SpeX spectroscopy.  The ratio of
the peak optical depth at 4.7\,$\mu$m ($\tau_{4.7}$) and the
standard deviation of the zero optical depth at the continuum is
2.6. The equivalent widths of the CO ice is estimated as
$\tau_{4.7} \cdot \sqrt{2\pi} \sigma_{4.7}$, where
$\sigma_{4.7}$ is the Gaussian sigma measured by fitting the
absorption profile. The equivalent width of CO ice is converted
to the column density using the integrated band strength
$A=1.1\times 10^{-17}$\,cm taken from
\cite{Gerakines:1995A&A...296..810G}. The results are shown in
Table~\ref{t1}.

The nearby template stars are significantly blueshifted from the
science targets, sometimes by as much as
$\sim$200\,km\,s$^{-1}$. A substantial part of the spectra on
the short-wavelength continuum is therefore lost at the
photospheric correction. With a tiny margin of continuum left,
it is hard to discern if the short-wavelength shoulder
4.655--4.665\,$\mu$m is part of the continuum or of the
absorption feature. If we assume this interval is part of the
continuum, the equivalent width of CO ice on source \#2
decreases to 2/3. None of the stars other than \#2 shows clear
CO ice absorption at 4.7\,$\mu$m. The upper limits of the
optical depth on those sources are 0.09--0.16, or
$\sim10^{17}$\,cm$^{-2}$ in the column density, with 3$\sigma$
significance.

The column of CO-ice detected on source \#2 is one of the smallest
records observed. The column density of CO with respect to water
ice is 16$^{+7}_{-6}$\,\%. If we take the narrow feature option
discussed above, the ratio is reduced to 11\,\%. This is smaller
than the typical fractional abundance of CO ice $\sim$30\,\%
found by {\em Spitzer} on background field stars toward nearby
star-forming regions \citep{Oberg:2011ApJ...740..109O}. The
small fractional abundance of CO ice at small column density is
qualitatively consistent with the larger $A_V$ threshold for CO
ice \citep[$A_V=6.0\pm4.1$; ][]{Chiar:1995ApJ...455..234C} than
that for water ice \citep[$A_V=3.2\pm0.1$;
][]{Whittet:2001ApJ...547..872W}.

\subsubsection{Comparison with laboratory spectra}

The  CO ice absorption detected toward source \#2 is compared to
the three laboratory spectra recorded by
\citet{Palumbo:1993A&A...269..568P} in Fig.~\ref{pa}. The
laboratory CO ice absorption is narrowest when deposited by
itself (2.5\,cm$^{-1}$ or 0.006\,$\mu$m in FWHM), and the
absorption maxima happens at the shortest wavelength
(2139\,cm$^{-1}$ or 4.675\,$\mu$m; top panel in
Fig.~\ref{pa}). The presence of co-adsorbate in general shifts
the line center longer (2137--2135\,cm$^{-1}$,
4.679--4.684\,$\mu$m) and broadens the line width
(3.5--24\,cm$^{-1}$ or 0.008--0.05\,$\mu$m). The effect is more
prominent for co-adsorbate with a larger dipole moment
\citep{Sandford:1988ApJ...329..498S,Ehrenfreund:1996A&A...315L.341E}.
Co-adsorbates, or matrix, sometimes add a satellite feature at
shorter wavelengths,  as in the case of water
\citep{Palumbo:1993A&A...269..568P}. Since it is hard to tell if
the shoulder of the observed depression at 4.655--4.665\,$\mu$m
belongs to the absorption or to the continuum, we have two
distinct solutions for the line widths. The narrow line solution
might well be reproduced by pure CO ice (top panel in
Fig.\ref{pa}), while the broader solution by the mixture of CO
and CO$_2$ (middle panel).  In both cases the line center of the
observed spectra (4.672--4.673\,$\mu$m) matches 
the laboratory spectra reasonably well.

The observed CO ice absorption does not seem to agree with the
laboratory spectra of CO ice mixed with H$_2$O ice (bottom
panel) where the absorption maximum happens at
4.677\,$\mu$m. We note that the observed CO ice spectrum is
normalized assuming that the intervals 4.645--4.655\,$\mu$m and
4.685--4.712\,$\mu$m are the continuum, which effectively
removes any excess absorption at those wavelengths. The mixture of CO and H$_2$O
 has prominent satellite feature at 4.647\,$\mu$m, which
is not fully covered in the astronomical spectrum. To see whether the
small margin of the observed continuum biases the matching of
the profiles, the laboratory spectrum of H$_2$O:CO is reduced in
the same way with the astronomical spectrum, assuming the same
intervals of the continuum. The line center of the H$_2$O:CO
laboratory spectrum changes little, and the conclusion stays the
same.

  \begin{figure}
     \center
     \includegraphics[width=0.45\textwidth]{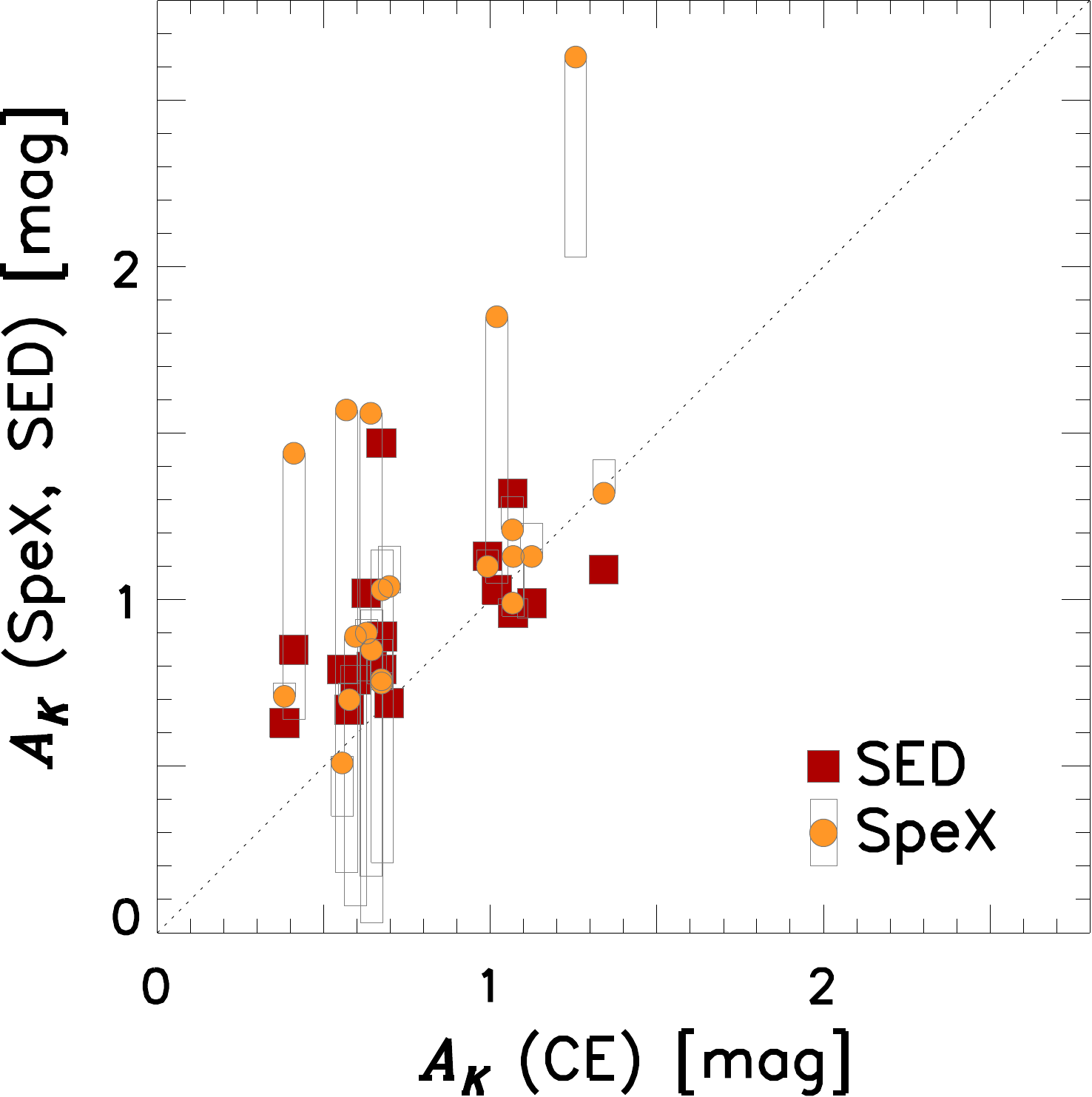}
      \caption{Comparison between $A_K$ from the near-infrared
        color excess (CE) technique by
        \cite{Roman-Zuniga:2010ApJ...725.2232R} against the SpeX
        templates and the SED fitting. The boxes attached to
        $A_K$ measured by SpeX templates delineate the range
        of extinctions with 3 best matching templates. The
        dotted line represents the case where the $A_K$'s would
        be identical to the color excess.}\label{ak1}
   \end{figure}

   \begin{figure*}
   \center
      \includegraphics[width=\textwidth, angle=0]{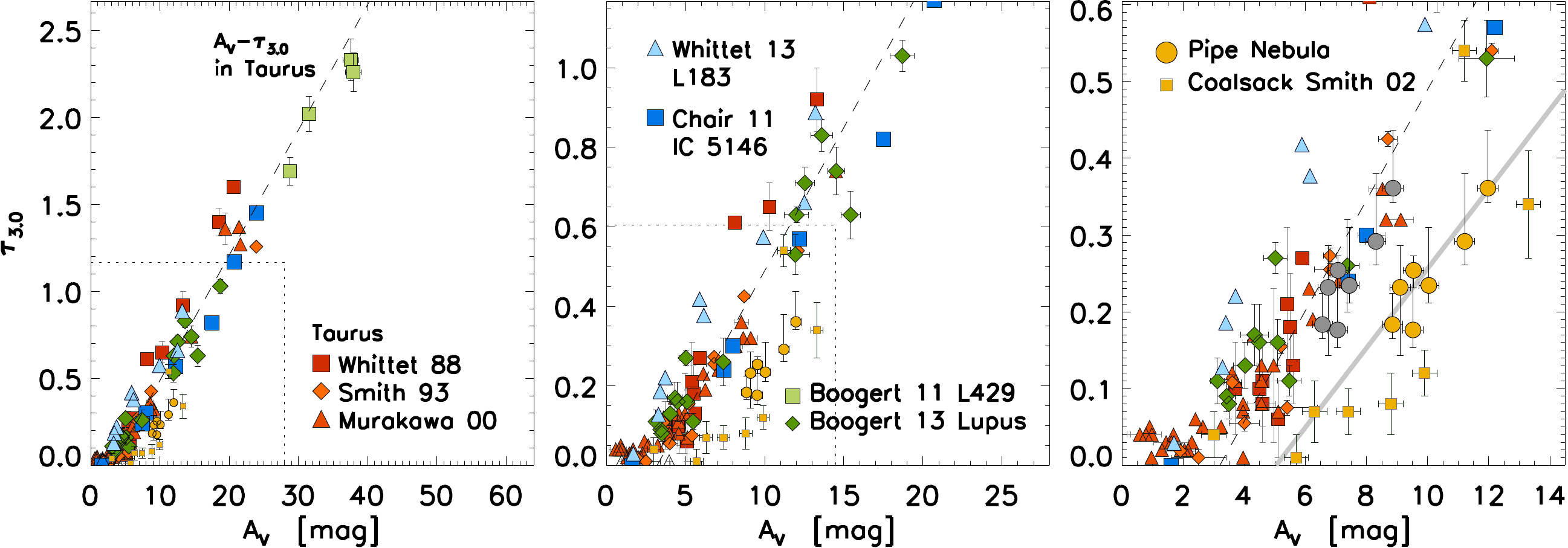}
      \caption{Peak optical depth of the water ice ($\tau_{\rm
          3.0}$) plotted against the visual extinction on the
        line of sight ($A_V$). The three panels are all
        identical, but the area shown is smaller going from left
        to right. The enlarged areas are shown by dotted lines
        in the left panels. The canonical relation between
        $\tau_{\rm 3.0}$ and $A_V$ in Taurus found by
        \cite{Whittet:2001ApJ...547..872W} is shown by dashed
        lines. The measurements toward the Pipe Nebula are shown
        by filled yellow circles. Other measurements shown in
        the panels are from star-forming regions such as Taurus
        \citep{Whittet:1988MNRAS.233..321W,Smith:1993MNRAS.263..749S,
          Murakawa:2000ApJS..128..603M}, Lupus
        \citep{Boogert:2013ApJ...777...73B}, and IC\,5146
        \citep[the Cocoon Nebula,][]{Chiar:2011ApJ...731....9C}.
        Two starless cores, L\,183
        \citep{Whittet:2013ApJ...774..102W}, L\,429
        \citep{Boogert:2011ApJ...729...92B}, and the globules in
        the Southern Coalsack \citep{Smith:2002MNRAS.330..837S}
        are shown as well. The linear regression fit to the data
        points from the Pipe Nebula is depicted by the gray
        line. Filled gray circles are the same as 
          yellow circles, but after reducing $A_V$ by 25\% from
        those in
        \citet{Roman-Zuniga:2010ApJ...725.2232R}. \label{w1}}
   \end{figure*}
\section{Discussion}
\subsection{Vistual extinction versus water ice optical depth\label{A_V}}
\subsubsection{Comparison of visual extinctions\label{extinction}}

There are at least three choices of extinctions to compare with the
ice optical depths: (1) the extinction  measured by the color
excess, which is presented in
\cite{Roman-Zuniga:2010ApJ...725.2232R} as an extinction map
     ($A_V$\,(CE) and $A_K$\,(CE) in Table~\ref{t1}); (2) the
     extinction  measured by the template match using the
     SpeX spectra ($A_K$(SpeX)); and (3) the extinction 
     measured by comparing the infrared photometry of the
     sources with the theoretical stellar models ($A_K$\,(SED)).

To obtain the extinction in method (3), the infrared photometry
of the objects from 1.2 to 22~$\mu$m are collected from 2MASS
and WISE catalogs. Theoretical stellar models and the SEDs are
computed by the PARSEC code \citep{Bressan:2012MNRAS.427..127B}
available through a web interface.\footnote{
  http://stev.oapd.inaf.it/cgi-bin/cmd\_2.7} The theoretical
SEDs are reddened by the same infrared extinction curve of
\cite{Boogert:2011ApJ...729...92B} used at the template matching
with the IRTF Spectral Library. The infrared extinction $A_K$
that best reproduces the observed SEDs are listed in
Table~\ref{t1} as $A_K$\,(SED).

The infrared extinctions on the lines of sight obtained by the
three techniques above are compared in Fig.~\ref{ak1}. The
visual extinction $A_V$ by
\cite{Roman-Zuniga:2010ApJ...725.2232R} is converted to $A_K$ by
multiplying $A_K/A_V=$0.112
\citep{Lombardi:2006A&A...454..781L}. In ideal circumstances,
all three extinctions should agree, which is apparently not the
case.

Each technique has its advantage and disadvantage. The color
excess technique does not measure the intrinsic color of the
individual stars, but uses the average color of the stars in the
nearby control field. The extinction is further averaged over a
20\arcsec\, square field; therefore, the technique is robust
against the outliers, but at the cost of not being sensitive to
individual lines of sight. Limited spatial resolution, compared
to a pencil beam measurement on an individual source, may miss
smaller structures with high visual extinctions.
\cite{Roman-Zuniga:2010ApJ...725.2232R} reported that the peak
$A_V$ in the densest part of the Pipe Nebula increases by a
factor of 4 in their extinction map; the spatial resolution is
three times higher than used by
\cite{Lombardi:2006A&A...454..781L}.

On the other hand, the critical drawback of the spectral
template and the SED technique is that the extinction is biased
to those sources in the Galactic Bulge far behind the Pipe
Nebula. \cite{Lombardi:2006A&A...454..781L} noted a clear
bifurcation of the stars toward the Pipe Nebula on the
near-infrared color-color diagram. The ``lower branch'' stars in
the diagram are brighter than those in the ``upper branch'', and
are likely a bulge population. The apparent brightness of the
lower branch stars peaks at $K\simeq$7\,mag, and closely matches
 our target selection. The line of sight to the Galactic
Bulge goes through the long distance behind the Pipe Nebula,
which is mostly diffuse, and not related to the ice absorption
inside the Pipe Nebula. The extinction biases toward the higher
$A_K$ by SpeX template (2) and SED (3) with respect to the
color-excess technique (1) can be seen in Fig~\ref{ak1}. We
therefore adopt the visual extinction obtained by the color
excess method to compare with the ice optical depths, despite the
risk of underestimating $A_V$. The uncertainty of $A_V$ in the
Shank is 0.33\,mag according to
\cite{Roman-Zuniga:2010ApJ...725.2232R}. The foreground
extinction toward the Pipe Nebula is estimated to be
$A_V=$0.12\,mag by \cite{Lombardi:2006A&A...454..781L} based on
the color excess of the Hipparcos stars with well-defined
distances, and is smaller than the uncertainty quoted above.

The peak optical depth of water ice $\tau_{\rm 3.0}$ is plotted
against $A_V$ in Fig.~\ref{w1}. A linear regression fit to the
seven sources where the water ice is positively detected renders
$m=0.050\pm0.026$ and the threshold extinction
$A_V^0=5.2\pm6.1$\,mag in $\tau_{3.0} = m \cdot (A_V -
A_V^0)$. The coefficients are not well constrained because of
the short baseline of $A_V$ and the large uncertainty in
$\tau_{3.0}$. Instead of going into a quantitative discussion on
$m$ and $A_V^0$, we would like to concentrate on the apparent
offset of $\tau_{3.0}$ from the canonical relation known in
Taurus \citep{Whittet:2001ApJ...547..872W}. The offset can be
interpreted in two ways: the visual extinction $A_V$ is too
large for the given $\tau_{3.0}$ in the Pipe Nebula, or the optical
depth of the water ice is too small for the given $A_V$. The first
possibility is discussed in the next sections.  We note that the
visual extinction  measured by the photometric color excess
adopted above is the most conservative option in this regard.

\subsubsection{Conversion of color excess to $A_V$}

Whatever  technique is used, the only observable we have to
constrain the line of sight extinction is the color of the
background sources measured either by photometric or
spectroscopic means. The observed color is compared with the
intrinsic color of the source, obtained either self-consistently
from spectroscopy or by an educated guess such as an average
color in a control field. The difference between the observed
and the intrinsic color amounts to the color excess $E(\lambda_1
- \lambda_2)$. The color excess is converted to the extinction
by $A_\lambda = c_\lambda\cdot E(\lambda_1 - \lambda_2)$, where
$c_\lambda$ is the conversion factor given by an extinction
law. Different authors adopt different extinction laws, thereby
a different conversion factor $c_\lambda$
\citep[e.g., ][]{Rieke:1985ApJ...288..618R,
  Cardelli:1989ApJ...345..245C,
  Indebetouw:2005ApJ...619..931I}. Some extinction laws commonly
used in the literature are constrained by the observations
through diffuse clouds, where the stars are visible at $B$ and
$V$. It may well be the case that the extinction law in diffuse
clouds differs from that in dense clouds because of the grain
growth and the altered grain compositions
\citep{Chiar:2007ApJ...666L..73C,Nishiyama:2009ApJ...696.1407N}.
In a dense cloud with visual extinction $A_V \gtrsim 10$\,mag,
few stars are visible, which makes it next to impossible to
calibrate an infrared extinction law against the optical
extinction to constrain $A_V/A_K$ empirically
\citep{Indebetouw:2005ApJ...619..931I}.

The extinction map of the Pipe Nebula by
\cite{Roman-Zuniga:2010ApJ...725.2232R}, used in the present
study, adopts the extinction law by
\citet{Indebetouw:2005ApJ...619..931I} to convert the color
excess to $A_K$, and \cite{Rieke:1985ApJ...288..618R} to convert
$A_K$ to $A_V$, which results in $A_V=6.4\,E(J-K)$ and
$A_V=16.2\,E(H-K)$. The visual extinctions used to constrain the
threshold $A_V$ in Taurus \citep{Whittet:2001ApJ...547..872W}
were collected from diverse sources. The color-extinction
conversions used are $A_V=5.3\,E(J-K)$ by
\cite{Whittet:1988MNRAS.233..321W}, $A_V=5.4\,E(J-K)$ by
\cite{Smith:1993MNRAS.263..749S}, $A_V=12\,E(H-K)$ by
\cite{Murakawa:2000ApJS..128..603M}, $A_V=4.6\,E(J-K)$ by
\cite{Elias:1978ApJ...224..857E} and
\cite{Tamura:1987MNRAS.224..413T}. The conversion factor we
adopt is indeed larger than those used in Taurus. If we reduce
a visual extinction by \cite{Roman-Zuniga:2010ApJ...725.2232R}
by a factor of 25\,\% to match 
\cite{Murakawa:2000ApJS..128..603M} which has the largest
difference, the apparent deviation of $\tau_{\rm 3.0}$ in the
Pipe Nebula from Taurus would almost disappear, even though all
measured $\tau_{\rm 3.0}$ still lie below the line expected in
Taurus (indicated by gray circles in Fig.~\ref{w1}).

This amount of uncertainty in the extinction law is entirely
conceivable. \cite{Nishiyama:2009ApJ...696.1407N} reconstructed
the wavelength dependency of $A_\lambda$/$A_{K_s}$ toward the
Galactic Center, using red clump stars in the Galactic
Bulge. The extinction ratio they found at $H$ and $K_s$ is
$A_H$/$A_{K_s}$=1.69, while
\cite{Roman-Zuniga:2010ApJ...725.2232R} adopted
$A_H$/$A_{K_s}$=1.55 from \cite{Indebetouw:2005ApJ...619..931I}
constrained by the 2MASS and Spitzer GLIMPSE surveys through the
Galactic plane toward $l=42\degr$ and $284\degr$. The smaller
$A_H$/$A_{K_s}$ of \cite{Roman-Zuniga:2010ApJ...725.2232R}
indeed increases $A_{K_s}$ by $\sim$25\,\% with respect to
\cite{Nishiyama:2009ApJ...696.1407N}, as $A_{K_s} =
\frac{1}{c-1}\, E(H-K_s)$, where $c=A_H/A_{K_s}$.

\subsubsection{Diffuse components}

Another way to account for the excess $A_V$ is by the
anisotropic distribution of diffuse components. The threshold
$A_V$ represents the presence of an outer skin where grains
remain bare. The bare grains contribute to the visual
extinction, but not to the optical depth of ices. We can think
of two geometries that increase the relative path length through
such a diffuse medium. For instance, the dense cloud could be
clumpy, and the line of sight could pass through more than one
dense region or dense core. Each dense core is covered by a
diffuse skin; therefore, the threshold $A_V$ would increase by
as much as the number of diffuse components that are along the
line of sight \citep{Smith:2002MNRAS.330..837S}. Alternatively,
the line of sight might also pass only a single dense core but
with various impact parameters
\citep{Murakawa:2000ApJS..128..603M}. Supposing a spherical
dense core embedded in a diffuse cloud and a line of sight that
barely grazes the surface of the dense core, the fraction of the
pathlength in the diffuse component would be larger than
when the sightline hits the very center of the core. The
contribution to $A_V$ from the diffuse component would therefore
larger. Such configurations are, however, not able to explain
the excess $A_V$ in the Pipe Nebula. The projected separation of
the targets we observed ranges from 0.08 to 0.63\,pc
(Fig.~\ref{p5}), and is similar to or larger than the typical
size of a single dense core ($\sim$0.1\,pc). If the particular
geometries of cloud cores are responsible for the extra visual
extinction, significant scatter is expected in the excess $A_V$
in Fig.~\ref{w1}. On the contrary, the optical depths of the
water ice line up almost in a straight line with small scatter
in $A_V$. A single constant offset in all $A_V$ prefers a common
diffuse screen that covers all lines of sight.

Based on the color excess of Hipparcos and Tycho stars,
\cite{Lombardi:2006A&A...454..781L} dismissed the presence of
such a diffuse component in front of the Pipe Nebula that
contributes more than $A_V=$1\,mag. As is discussed in \S
\ref{extinction}, the targets we observed have a bias of being
in the Galactic Bulge. A diffuse screen far behind the Pipe
Nebula is, however, also not likely.  Source \#2 shows the
largest $\tau_{\rm 3.0}$, and has the largest offset in $A_V$
from the canonical $\tau_{\rm 3.0}$ relation in Taurus. The
infrared extinction $A_K$ toward \#2 obtained by the spectral
template and the SED analysis agree well with the color excess,
suggesting \#2 is immediately behind the Pipe Nebula. If an
extra diffuse component is responsible for the enhanced visual
extinction, the screen is not far away from the Pipe Nebula, but
could be part of it.

   \begin{figure*}
     \center     
      \includegraphics[width=0.95\textwidth]{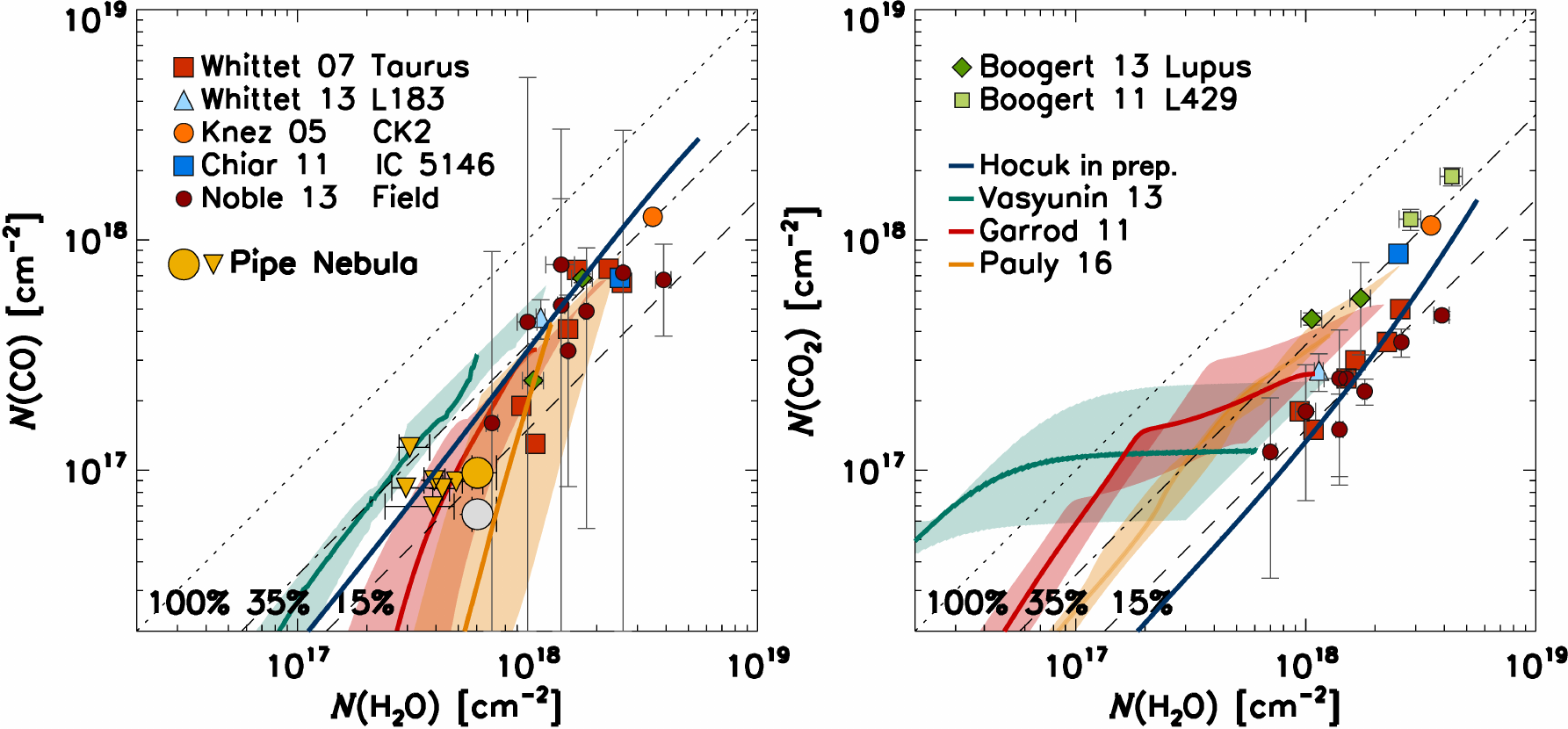}
      \caption{Column densities of CO (left) and CO$_2$ (right)
        ices plotted against that of water ice. The measurements
        of CO ice toward the Pipe Nebula are all upper limits
        (yellow triangles) except for 17312818$-$2631268 (source
        \# 2) shown in the yellow circle. The gray circle is the
        column density of CO ice on source \#2 as well, but
        measured assuming that the short-wavelength shoulder
        4.655--4.665\,$\mu$m is part of the continuum (see
        \S\,\ref{t47}). No observation of CO$_2$ ice is
        available in the Pipe Nebula. Other measurements shown
        are from Taurus \citep{Whittet:2007ApJ...655..332W},
        L\,183 \citep{Whittet:2013ApJ...774..102W}, CK\,2
        \citep{Knez:2005ApJ...635L.145K}, IC\,5146
        \citep[Q21-1;][]{Chiar:2011ApJ...731....9C}, assorted
        quiescent clouds \citep{Noble:2013ApJ...775...85N},
        Lupus \citep{Boogert:2013ApJ...777...73B}, and L\,429
        \citep{Boogert:2011ApJ...729...92B}. All the
        measurements are against the continua of background
        field stars through quiescent clouds. Only those lines
        of sight where both CO and CO$_2$ ices are positively
        detected are shown. Four simulations of ice formation
        are compared with the observations \citep[Hocuk et
          al. in prep.; ][]{Vasyunin:2013ApJ...762...86V,
          Garrod:2011ApJ...735...15G,Pauly:2016ApJ...817..146P}.
        The number of layers presented in the simulations are
        converted to the column density by assuming that one
        monolayer corresponds to $9.6\times10^{15}$\,cm$^{-2}$
        (Appendix~\ref{ap2}).  Model column
          densities calculated with conversion factors that are two
          times larger and smaller are shown by the shaded
          areas.\label{n3}}

   \end{figure*}
\subsection{Radiation field \label{rad}}

The second interpretation of Fig.~\ref{w1} is that the optical
depth of the water ice in the Pipe Nebula is too small for a given
$A_V$, possibly because of the elevated interstellar radiation
impinging on the Pipe Nebula. In a quiescent cloud,
photoprocesses (dissociation and desorption) are the primary
hindrances to  ice mantle formation.
\cite{Hollenbach:2009ApJ...690.1497H} gives a succinct formula
of the threshold $A_V$ for ice formation that is proportional to
$\ln \left(G_0/n\right)$, where $n$ is the gas number density
and $G_0$ is the far-UV radiation field outside the cloud in 
units of Habing field \citep{Habing:1968BAN....19..421H}. A
stronger radiation field implies faster photodissociation and
desorption, while the higher density accelerates the accretion
of molecules on the grain surface.
Using the formula (17) of \cite{Hollenbach:2009ApJ...690.1497H},
the radiation field must be $\sim 18\,G_0$ to  double the
threshold $A_V$ from 1.6\,mag (Taurus) to 3.2\,mag.

The Pipe Nebula is close to $\rho$~Oph in projection, where a
large $G_0$ is inferred by the presence of high-mass stars in
the nearby Sco-Cen OB association
\citep{Wilking:2008hsf2.book..351W}. The water ice in $\rho$~Oph
shows a similar trend to the Pipe Nebula with smaller optical
depths than in Taurus at the same visual extinctions
\citep{Tanaka:1990ApJ...352..724T,Chiar:1995ApJ...455..234C}. The
small column density of CO ice with respect to water ice in
$\rho$~Oph \citep[$\sim$20\,\%;][]{Kerr:1993MNRAS.262.1047K} may
corroborate that the Pipe Nebula and $\rho$~Oph are bathed by
stronger radiation field than other nearby dark clouds.

A stronger radiation field raises the dust temperature in a
cloud
\citep[e.g.,][]{Hollenbach:1991ApJ...377..192H,Galli:2002A&A...394..275G}.
The dust temperatures measured in the Pipe Nebula by
far-infrared imaging with Herschel are 15--19\,K at
$A_V$=10\,mag \citep{Forbrich:2014A&A...568A..27F}. This is
higher than other similar quiescent clouds such as L\,1595 and
B\,68 \citep[10--13\,K; compilation presented
  in][]{Hocuk:2017arXiv170402763H}. In order to raise the dust
temperature to 15--19\,K at the depth $A_V=10$\,mag, the
external $G_0$ must be in the range of 25--100, according to the
prescription given by \citet{Hocuk:2017arXiv170402763H}.

Little is known about $G_0$ around the Pipe Nebula by non-dust
UV probes. \citet{Kamegai:2003ApJ...589..378K} compare their
[CI] observation in $\rho$ Oph with the PDR model by
\citet{Kaufman:1999ApJ...527..795K}, and estimated $G_0$ to be
$\sim100$. On the other hand, \citet{Bergin:2006ApJ...645..369B}
argues that $G_0$ must be 0.2 or smaller to reproduce C$^{18}$O
line emission at the outer part of B\,68. B\,68 is located
between the Pipe Nebula and $\rho$ Oph, but much closer to the
former in the projection on the sky.

\subsection{Ice mantle in formation \label{chemevo}}

Another possible explanation of the small optical depth of water
ice in the Pipe Nebula is that the  formation of the ice mantle has just
started. The optical depths of the water ice in nearby
star-forming regions, Taurus, Lupus, and IC\,5146 (the Cocoon
Nebula), and in the starless cores L\,183, L\,429, and the
globules in the Southern Coalsack, are plotted in Fig.~\ref{w1}
along with the present observation in the Pipe Nebula. This is
not an exhaustive survey, but the recent spectroscopy by the
{\em Spitzer} Space Telescope are selected with priority. The
optical depth of the water ice in the Pipe Nebula is lower than
that in Taurus at the same $A_V$
\citep{Whittet:1988MNRAS.233..321W,Smith:1993MNRAS.263..749S,
  Murakawa:2000ApJS..128..603M}, although many sightlines in
Taurus are also in quiescent parts of the cloud. The optical
depth of water ice in Lupus \citep{Boogert:2013ApJ...777...73B}
and IC\,5146 \citep{Chiar:2011ApJ...731....9C} aligns well with
Taurus. The good correlation between $A_V$ and $\tau_{\rm 3.0}$
in star-forming regions implies that these environments are old
enough for the growth of the ice mantle to have reached steady
state. The ice column density increases  proportional to
$A_V$, implying that this is due to the accumulation of the mass
along the line of sight, rather than to the formation of more ice
mantles.

L\,183 \citep{Whittet:2013ApJ...774..102W} and L\,429
\citep{Boogert:2011ApJ...729...92B} are evolved starless cores
\citep[prestellar cores,][]{Crapsi:2005ApJ...619..379C} with no
active star formation spotted currently. The water ice optical
depths in L\,183 and L\,429 are not too different from those in
Taurus. The visual extinction toward L\,183 at the far-infrared
continuum peak reaches $\sim$150\,mag
\citep{Pagani:2004A&A...417..605P}. L\,429 is seen in absorption
against background continuum emission up to the wavelength
70\,$\mu$m by MIPS on {\em Spitzer}
\citep{Stutz:2009ApJ...690L..35S}. The visual extinction at the
core center reaches 35--130\,mag. The large visual extinction
implies that both starless cores are in a physically evolved
phase and on the verge of starting star formation.

On the contrary the largest visual extinction in the Southern
Coalsack found by the near-infrared color excess technique is
$A_V$$\sim$12\,mag \citep{Lada:2004ApJ...610..303L}, which is
similar to or less than the maximum extinction in the Shank of
the Pipe Nebula
\citep[$A_V\sim$15\,mag;][]{Roman-Zuniga:2010ApJ...725.2232R}.
\cite{Smith:2002MNRAS.330..837S} reported that the optical depth
of water ice is significantly lower than Taurus at the same
$A_V$, similar to the Pipe Nebula, but in a more extreme manner
(right panel of Fig.~\ref{w1}). All alternative explanations
discussed in \S\,\ref{A_V} also apply to the Southern Coalsack;
however, the correction in $A_V$ required to align the
observation in the Southern Coalsack to Taurus is $\sim$5\,mag,
or $\sim$50\,\% reduction of $A_V$, which is not easy to justify
even with typical uncertainties in the extinction laws. It is
intriguing that among the many sightlines observed, the two
clouds that are possibly at the earliest stage of evolution show
the lowest $\tau_{\rm 3.0}$.
\subsection{Comparison with simulations\label{simcom}}

We discuss in this section how the observed column densities of
water, CO, and CO$_2$ ice fit in with the latest simulations. The
observations on the field stars behind nearby star-forming
regions are added from the literature, and shown together with
the simulations in Fig.~\ref{n3}. Since the formation of the
three most abundant ices--water, CO, and CO$_2$--are tightly
correlated to each other, the column densities of CO$_2$ is also
included. The present observations in the Pipe Nebula sample the
lowest end of the column densities of water and CO ice.

The comparison with the simulations should be taken with
caution. First, the observed column densities of ice may
increase in either way, when more mantles build up on the grain
surface or when the gas in the cloud aggregates on the line of
sight. Even after the molecules in the gas are completely
depleted, column densities of ice could still grow, if the cloud
continues to accrete material. The larger column densities of
ice in Taurus  compared to the Pipe Nebula possibly represent
thicker ice mantles in Taurus. However, if we compare a few lines
of sight of different $A_V$ within Taurus, the larger column
density than the other may simply reflect more material
accumulated on the line of sight rather than the growth of the
ice mantles on the grains. 

We compare four different simulations with the observations in
quiescent clouds. The chemical composition of ice computed by
\citet{Vasyunin:2013ApJ...762...86V},
\citet{Garrod:2011ApJ...735...15G}, and
\citet{Pauly:2016ApJ...817..146P} are along evolutionary
tracks. In these models, the clouds are embedded in a radiation
field, and dynamically collapse one-dimensionally in free fall.
The chemical compositions are calculated for $A_V$ and the dust
temperature given by the dynamical models. Hocuk et al. in
prep., which is a follow-up of the work
\citet{Hocuk:2015A&A...576A..49H}, is instead a
three-dimensional simulation that gives a snapshot of a cloud at
a given time. The range of $N$(H$_2$O) represents different
parts of the cloud with different degrees of material
aggregation, as opposed to the other simulations that provide
temporal variations. The model by Hocuk et al. in prep. is
best compared to the observations performed in a single cloud or
cloud core.

Second, \citet{Vasyunin:2013ApJ...762...86V},
\citet{Garrod:2011ApJ...735...15G}, and
\citet{Pauly:2016ApJ...817..146P} do not provide the composition
of ices in column densities, but the number of layers of the ice
mantles. We assume a monolayer corresponds to
$9.6\times10^{15}$\,cm$^{-2}$ to convert the number of layers to
column densities (see Appendix \ref{ap2}). Such simple
conversion has to be taken with caution because the conversion
factor is calculated with constant gas density ($n_{\rm
  H}=10^4$\,cm$^{-3}$ in the present case) during the dynamical
collapse of the cloud, which is obviously not the case. One
should bear in mind that the column densities shown in
Fig.~\ref{n3} from these models are simply proportional to the
number of layers on a single grain. The water ice column
observed on \#2 amounts to 50--80 layers, assuming the
layer-column conversion factor.
In the case of CO$_2$ ice in
\citet{Vasyunin:2013ApJ...762...86V},
\citet{Garrod:2011ApJ...735...15G}, and
\citet{Pauly:2016ApJ...817..146P}, where there are few overlaps
with the observations, $N$(CO$_2$)/$N$(H$_2$O) ratios at the end
points of the simulations are the critical figures to compare
with the observations, as the column densities continue to grow
from there, linearly, as the cloud collapses further.

We look at each simulation closely below. Hocuk et al. in
prep. solve classical rate equations to calculate the
abundances of molecules in gas and on grains. Major surface
processes such as photo- and chemical desorption,
photodissociation, surface reactions through thermal diffusion,
and tunneling are fully taken into account, as they are in the
other simulations discussed here. The magnetohydrodynamical
evolution of a cloud is numerically traced with time-dependent
chemistry. The simulation, however, produces slightly less
CO$_2$ ice than the observations because the adopted dust
temperature is lower than in the other simulations, which
slows down the diffusion of CO on the surface to form CO$_2$ by
the reaction between CO and OH.

\cite{Vasyunin:2013ApJ...762...86V} use the macroscopic Monte Carlo
technique to follow the formation of molecules on a grain
surface in a stochastic manner.  Layer-by-layer treatment allows
only the molecules in the outermost layers to participate in the
chemical reactions, while those deep inside the mantles are kept
inert. \cite{Vasyunin:2013ApJ...762...86V} reproduce CO ice
abundance moderately well with a hint of overproduction, in
particular in comparison with the small column density of CO ice
on \#2. In their model, CO$_2$ forms in competition with
water. As long as H$_2$O continuously photodissociates in the
gas and on the surface, the ample supply of the OH radical is
available for the surface reaction CO + OH. As the cloud
evolves, $A_V$ becomes larger and water does not
photodissociate any more; CO$_2$ formation almost completely stops
because OH is locked in H$_2$O. The transition is seen in the
flattening of the CO$_2$ ice at $\sim 10^{17}$\,cm$^{-2}$, where
CO ice starts building up on the surface. Here the cue of the
switchover is  $A_V$, which halts the photodissociation of the
water.

\cite{Garrod:2011ApJ...735...15G} solve rate equations to
calculate the surface molecular abundances, but with the
modified reaction rates by \cite{Garrod:2008ApJ...682..283G} to
better represent the stochastic cases where the limited number
of molecules are on a single grain surface. The surface
chemistry is calculated in three phases, gas, surface, and
mantle, and the molecules in the mantle are treated as
chemically inert. The chemical and dynamical models of
\cite{Pauly:2016ApJ...817..146P} are the same as
\cite{Garrod:2011ApJ...735...15G}, but the former emphasizes the
importance of the grain size to the surface chemistry 
because even a few degrees of change in the dust temperature has
a large impact on the chemistry, and smaller grains are warmer
if embedded in the same radiation field. In the case of
\cite{Garrod:2011ApJ...735...15G} and
\cite{Pauly:2016ApJ...817..146P}, the dust temperature is the
cue that stalls the active CO$_2$ formation. While OH radicals are
pinned on the grain surface and immobile, CO is mobile as long
as the dust temperature is higher than 12\,K. Every CO molecule on the
surface is eventually converted to CO$_2$ by the reaction with
OH. As the cloud continues to evolve, and the dust grains cool
down, the thermal diffusion of CO stops. CO$_2$ formation slows
down, and takes place only when  atomic oxygen is trapped in
the same site with CO, and reacts with incoming atomic hydrogen
to form an OH radical on the spot
\citep{Goumans:2008MNRAS.384.1158G}. Perceptible amount of CO
starts being left on the surface only at this stage, as is seen
in $N$(CO) rising at the turnaround of $N$(CO$_2$) in
\cite{Garrod:2011ApJ...735...15G}.

The three simulations of ice composition along the evolutionary
tracks
\citep{Vasyunin:2013ApJ...762...86V,Garrod:2011ApJ...735...15G,Pauly:2016ApJ...817..146P}
stop when a significant fraction of the molecules in the gas
phase are depleted after building 150--300 layers of ice
mantles, and the net deposition rate decreases to zero. The
$N$(CO$_2$)/$N$(H$_2$O) ratios end up at 20--30 \%, and are
consistent with the observations in the literature.
\cite{Garrod:2011ApJ...735...15G} and
\cite{Pauly:2016ApJ...817..146P} show a smaller CO-ice over
water-ice ratio at the early phase of the ice mantle formation,
which is in agreement with our observation toward \#2.  We note
that from observations of CO and CO$_2$ ice, in particular the
latter, in a cloud that are still relatively diffuse,
[$N$(H$_2$O) $\approx 10^{17}$\,cm$^{-2}$] will provide vital
clues to critically discriminate among the models discussed
above.

Yet another way to form CO and CO$_2$ ice on a grain surface,
but not taken into account in the models above, is through an
energetic path. \cite{Mennella:2004ApJ...615.1073M} found in the
laboratory that CO and CO$_2$ ice are newly formed on the
synthesized hydrocarbon grain covered by water ice after
applying a 30\,keV He$^+$ ion beam. The experiments are meant to
simulate the cosmic ray impact on dust grains in the
interstellar medium. Here, the formation of CO$_2$ ice does not
require the pre-deposition of CO on the substrate, but the
supply of atomic carbon comes from the grain surface covered by
water ice but eroded by the ion irradiation. The ion irradiation
produces about the same amounts of CO and CO$_2$ ices
\citep{Mennella:2004ApJ...615.1073M}, and is consistent with the
observations. The simultaneous production of CO and CO$_2$ is in
contrast to the classical diffusion reaction pathway, such as in
the simulations discussed in Fig.~\ref{n3}, where only one of
the two molecules actively forms at a given moment.

\section{Conclusion\label{summary}}

We started a spectroscopic survey of ices in the Pipe Nebula to
measure the pristine compositions of molecules in ice mantles
that have not been altered by feedback from star formations.
The summary of the observations is as follows:

\begin{enumerate}

\item We obtained 1.9--5.3\,$\mu$m spectra of background field
  stars behind the Pipe Nebula with SpeX spectrograph at the
  IRTF. Water ice absorption was detected at 3.0\,$\mu$m on seven
  lines of sight. The peak optical depths of the water ice are
  about half as large as those on the sources in Taurus with
  similar visual extinctions.

\item Possible explanations of the difference are (i) the visual
  extinction through the Pipe Nebula is overestimated; (ii) the
  radiation field impinging on the Pipe Nebula is larger than
  that on the Taurus Molecular Cloud; or (iii) the formation of
  the ice mantle has just started, that is, the Pipe Nebula is
  in an earlier phase of ice evolution than Taurus is.

\item The seven sources on which water ice was detected are
  further investigated with the NIRSPEC spectrograph at Keck
  II. The source with the highest water ice optical depth shows
  CO ice absorption at 4.7\,$\mu$m. The ratio of the CO ice
  column density to the water ice is 16\,\% and smaller than the
  $\sim$30\,\% often observed in nearby star-forming regions.

\item The  column densities of water, CO, and CO$_2$ ice collected
  from the literature and from the present observations
  are compared to recent simulations of ice formation on
  grains. Some simulations predict low CO ice abundance with
  respect to water ice at the early phase of the ice mantle
  formation, which is consistent with what we found in the Pipe
  Nebula in one source. Future observations of ices in diffuse,
  less evolved clouds are the key to understanding the
  elemental surface processes and the evolution of ice mantles
  in molecular clouds.

 \end{enumerate}

\begin{acknowledgements}
  We appreciate the constructive feedback of the anonymous
  reviewer of this paper. We thank all the staff and crew of the
  IRTF for their valuable assistance in obtaining the data. 
    The data presented here were obtained at the W.M. Keck
    Observatory, which is operated as a scientific partnership
    among the California Institute of Technology, the University
    of California, and the National Aeronautics and Space
    Administration.  The Observatory was made possible by the
    generous financial support of the W.M. Keck Foundation.  We
  appreciate the hospitality of the Hawaiian community that made
  the research presented here possible. The authors sincerely
  thank Anton Vasyunin who taught us about his models and how to
  convert a monolayer to a column density. We also thank
  Wing-Fai Thi, Felipe Alves de Oliveira, Carlos
  Rom\'an-Z\'u\~niga, and Gabriel Pellegatti Franco for their
  helpful discussion on the uncertainty of the extinction toward
  the Pipe Nebula and general. This research has made use of the
  SIMBAD database, operated at CDS, Strasbourg, France.  The
  publication makes use of data products from the Two Micron All
  Sky Survey, which is a joint project of the University of
  Massachusetts and the Infrared Processing and Analysis
  Center/California Institute of Technology, funded by the
  National Aeronautics and Space Administration and the National
  Science Foundation.  The publication makes use of data
  products from the Wide-field Infrared Survey Explorer, which
  is a joint project of the University of California, Los
  Angeles, and the Jet Propulsion Laboratory/California
  Institute of Technology, funded by the National Aeronautics
  and Space Administration.   M.G. is supported by the German
    Research Foundation (DFG) grant GO 1927/6-1. P.C., S.C., and
    M.S. acknowledge the financial support from the European
    Research Council (ERC, Advanced Grant PALs 320620).
\end{acknowledgements}
\bibliographystyle{aa} 
\bibliography{aa}
\begin{appendix}
  \section{SpeX/IRTF spectra  \label{ap1}}

Infrared spectra of 21 sources obtained by SpeX are presented in
Fig.~\ref{s9}. Those matching  a spectral type earlier
than M7 are shown here. The spectra are ordered from top to
bottom according to the visual extinction measured by
\citet{Roman-Zuniga:2010ApJ...725.2232R} with the highest $A_V$
first. The wavelength interval near 3.3\,$\mu$m is removed from
the presentation because the atmosphere is totally opaque due
to the telluric methane absorption. The optical depth of water
ice is calculated with respect to the matching template spectra
$f_0$ as $\tau = -\ln\frac{f}{f_0}$.

 \begin{figure*}
      \center
      \includegraphics[width=0.88\textwidth]{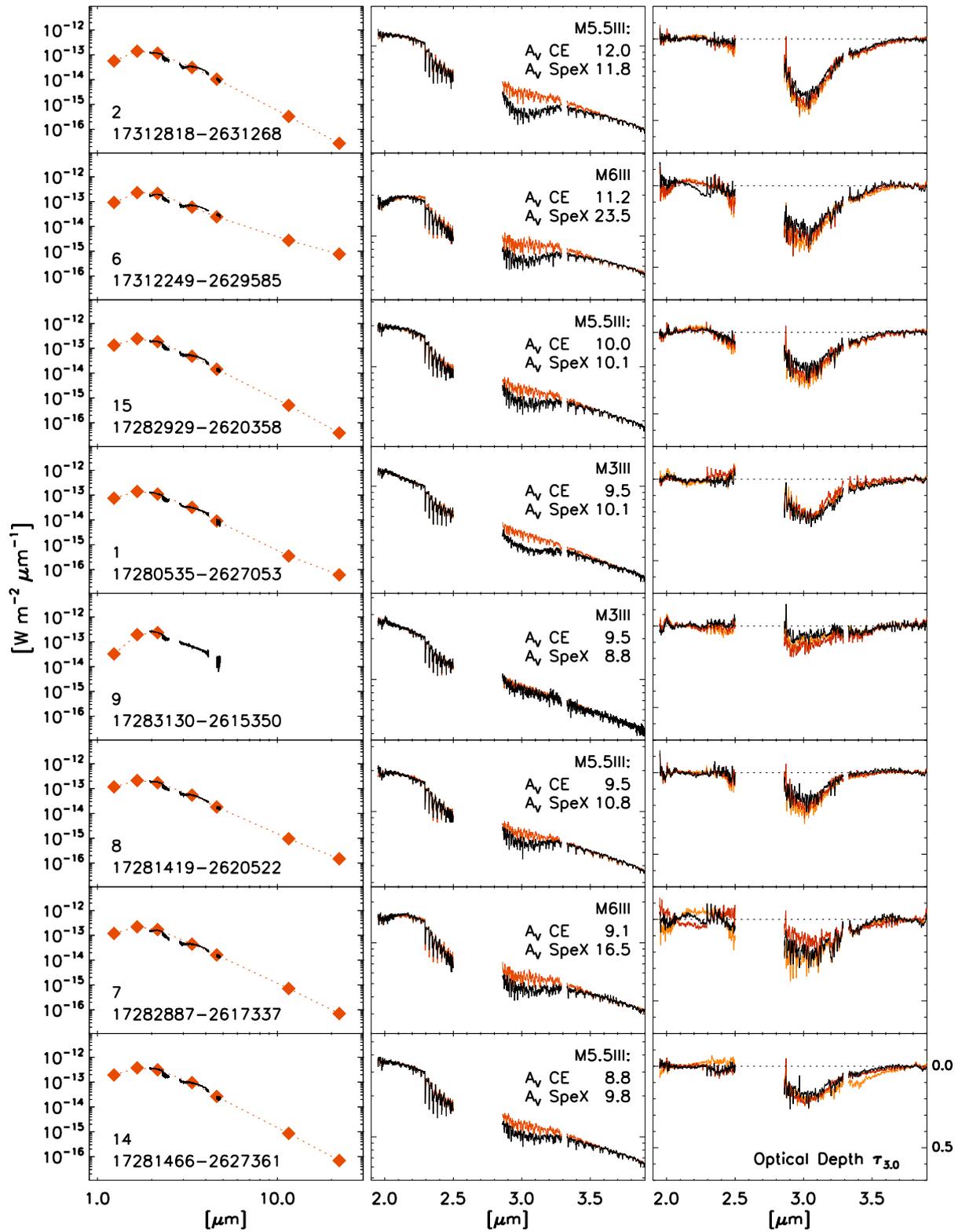}
      \caption{Left:  SpeX spectra overlaid with 2MASS
        and WISE photometry. The identification of the sources
        in Table~\ref{t1} and Figure~\ref{p5}, and the 2MASS
        names are given at the bottom left. The SpeX spectra are
        scaled to $K_s$ (2MASS) and $W1$ (WISE) band photometry,
        but by multiplying a single factor without adjusting $K$
        and $L$ band spectra. Middle:  Zoom of the
        spectra from 2.0 to 3.9\,$\mu$m. The best match template
        spectrum from IRTF spectral library is shown in orange
        in background. The excess absorption centered at
        3.0\,$\mu$m is attributed to the water ice. The best
        match spectral type, the visual extinction as measured
        by \citet{Roman-Zuniga:2010ApJ...725.2232R} (``CE'' for
        color excess technique), and by matching the
        template stars (``SpeX'') are shown at the top
        right. Right: Optical depth spectra at
        3\,$\mu$m.  The optical depth spectrum calculated with
        the best matching template is shown in black, the second
         best in red, and the third best in orange to highlight
        the systematic uncertainty that stems from the choice of
        the templates.}
      \label{s9}
\end{figure*}
\setcounter{figure}{0}
\begin{figure*}
       \center
      \includegraphics[width=0.98\textwidth]{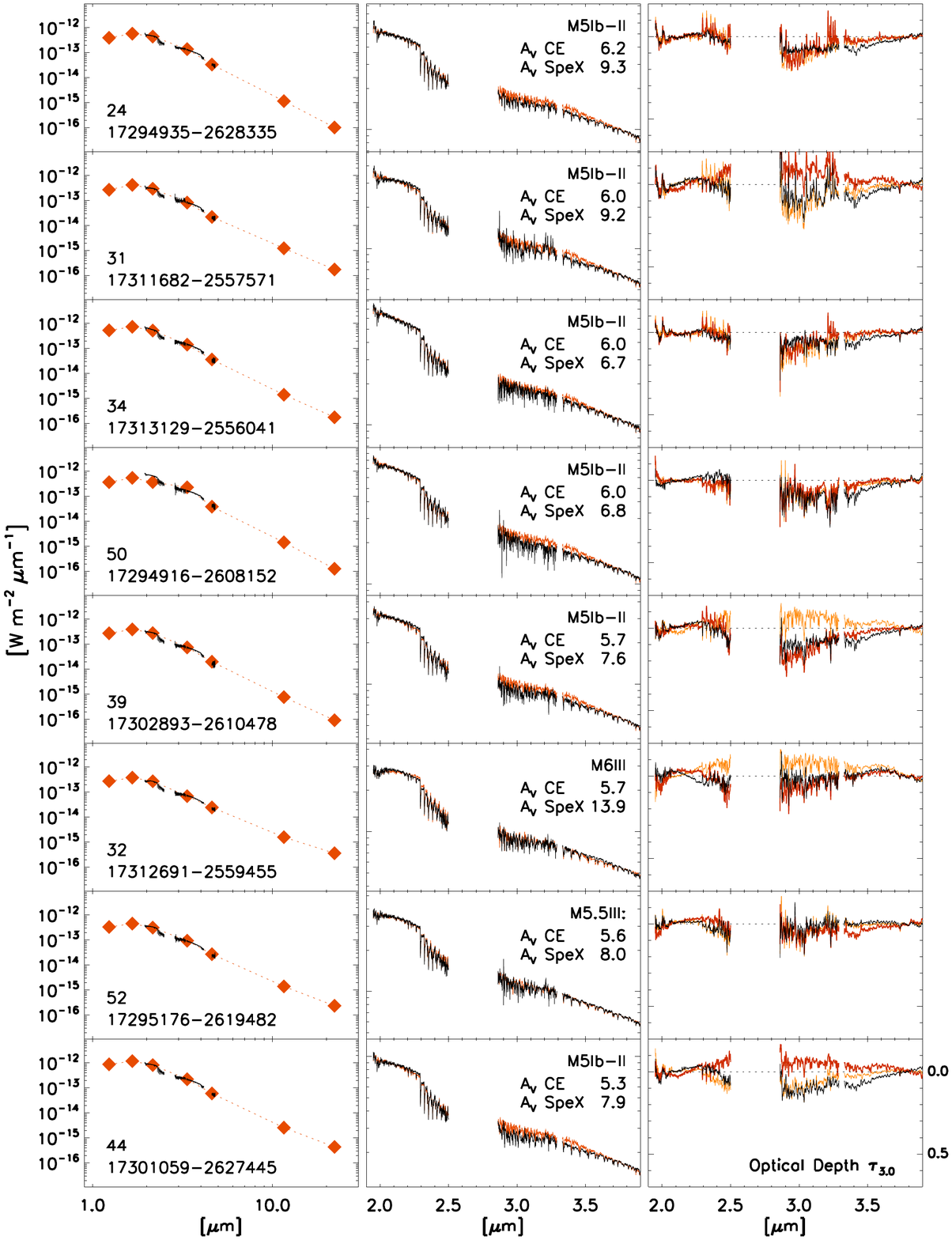}
       \caption{Continued.}
  \end{figure*}
 \setcounter{figure}{0}
 \begin{figure*}
       \center
       \includegraphics[width=0.98\textwidth]{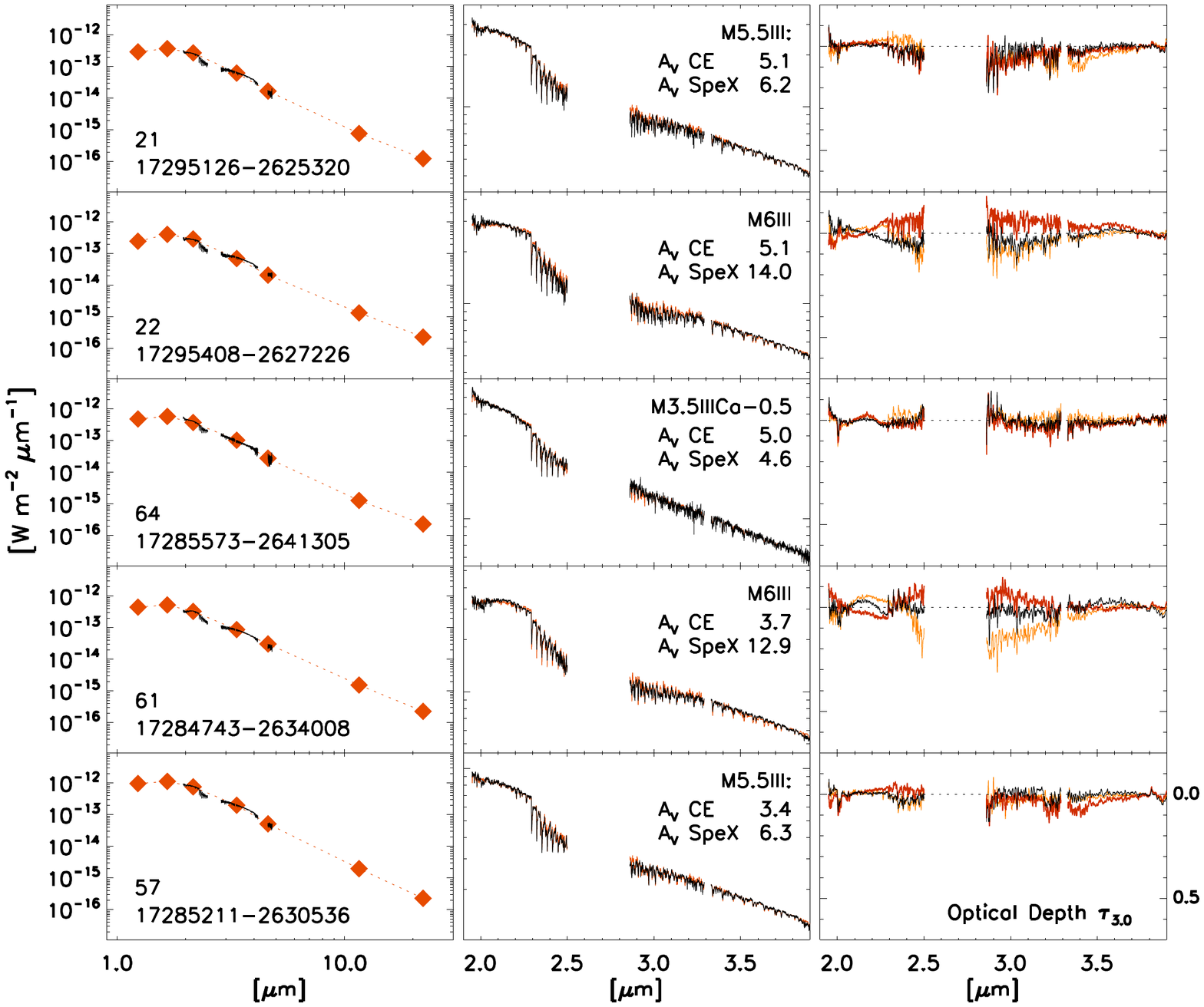}
       \caption{Continued.}
 \end{figure*}

\clearpage
   \begin{figure}[h]
     \center
      \includegraphics[width=0.45\textwidth]{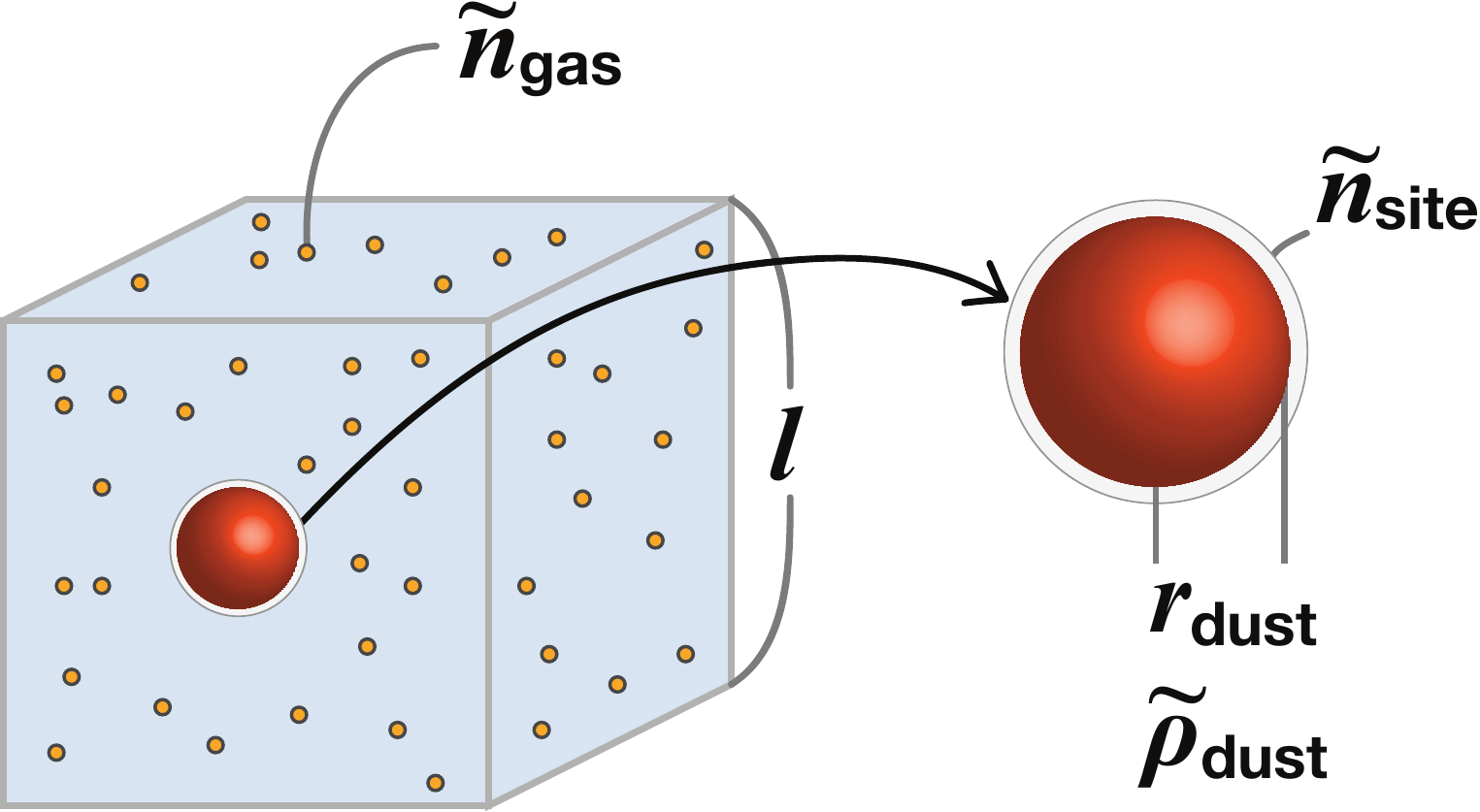}
      \caption{How the number of monolayers are translated to
        the column density of molecules. The first step is
        calculating the size of the cube ($l$) assigned to a
        single grain shown as a red ball in the picture.
  \label{grain}}
   \end{figure}
\section{Translation of monolayers to column density \label{ap2}}

   The simulations discussed in the paper, except Hocuk et
   al. in prep., do not provide the expected amount of ice
   column density that we can directly compare with the
   observations. Instead, they assume a surface of a single
   grain, and solve time-dependent chemistry with the physical
   conditions given locally, such as gas density, $A_V$ counted
   from the edge of the cloud, and dust temperature. The output
   is the number of layers of particular molecules in ice
   stacked on the grain surface. The simulations are not bound
   to any macroscopic setup, e.g., the size of the cloud. In
   order to compare the simulation with the observations, we
   need to translate the simulated ice molecular compositions on
   the grain surface to the column density of the ices through
   the cloud cores.


We start with a single spherical grain with the radius $r_{\rm
  dust}=0.1$~$\mu$m and calculate the size of the cube assigned
to this particle that does not overlap with neighboring
particles (Fig.~\ref{grain}).  The size of the box $l$ is
estimated, assuming a typical gas-to-mass ratio. The mass of a
single grain particle is
\[
{\tilde{M}_{\rm dust}}
= \frac{4}{3} \pi r_{\rm dust}^3 ~\tilde{\rho}_{\rm dust} {\rm ~~~[g],} 
\]
\noindent
where  $\tilde{\rho}_{\rm dust}$ is the mass density of the dust grain. The
total number of  gas-phase particles in the box, i.e., the number of 
mean mass molecules, $\tilde{n}_{\rm gas}$ amounts to

\begin{eqnarray*}
\tilde{n}_{\rm gas} &=& \frac{\tilde{M}_{\rm gas}}{\mu_m m_H} \\
             &=& \frac{\tilde{M}_{\rm dust}}{\mu_m m_H} \cdot \frac{\rho_{\rm gas}}{\rho_{\rm dust}}\\
             &=&  4.5 \times 10^{11},
\end{eqnarray*}

\noindent
where $\tilde{M}_{\rm gas}$ is the total mass of the gas in the
box, $\mu_m$ is the mean molecular weight 1.4, and $m_H$ is the
mass of a hydrogen atom $1.67 \times 10^{-24}$~g. We used
  gas-to-dust mass ratio $\rho_{\rm gas}/\rho_{\rm dust}\sim
100$, and the mass density of the grain $\tilde{\rho}_{\rm
  dust}= 2.5$~g\,cm$^{-3}$. The typical transversal dimension
of the cores in the Pipe Nebula, as measured by
\citet{Roman-Zuniga:2010ApJ...725.2232R}, is $\sim$0.1~pc in
Shank and the peak visual extinctions through them are
$A_V$$\sim$10\,mag. The number densities of the gas in the
cores are on the order of $n_{\rm H}\approx10^4$~cm$^{-3}$. This
is consistent with the positive detection of C$^{18}$O by
\cite{Onishi:1999PASJ...51..871O} in these cores as well.  The
size of the cubic box is then

\begin{eqnarray*}
 l &=& \sqrt[\leftroot{-1}\uproot{2}\scriptstyle 3]
        {\frac{\tilde{n}_{\rm gas}}{n_{\rm H}}}\\
   &=&  3.5 \times 10^2 {\rm ~~~[cm].}
\end{eqnarray*}

\noindent
On the other hand, the number of sites in one monolayer
  that molecules can stick to is

\begin{eqnarray*}
  \tilde{n}_{\rm site} &=& 4 \pi r_{\rm dust} ^2 / s^2 \\
                &=& 1.4 \times 10^6 {\rm ~~~[molecule~mly^{-1}]}, 
\end{eqnarray*}

\noindent
where $s$ is the scale of the separation between the molecules
on the surface, and therefore $s^2$ is the area that one molecule
holds. We assume that the grain surface is fully covered by a
single layer of ice, and $s=3$\,$\AA$. Scaling the result to a unit volume cm$^{-3}$, the
number density of the molecules when a grain is fully covered by
a single layer of ice is

\begin{eqnarray*}
   n_{\rm site} &=& \tilde{n}_{\rm site} /l^3 \\
     &=& 3.1 \times 10^{-2} {\rm ~~~[molecule~mly^{-1}\,cm^{-3}]}.
\end{eqnarray*}

\noindent
The translational dimension of the cores in the Pipe Nebula is again
$L\sim$0.1 pc or $3.1 \times 10^{17}$~cm. Therefore, the monolayer
of molecular ice in our particular case translates to a column
density of
\begin{eqnarray*}
  N_{\rm mly} &=& n_{\rm site} \, L \\
              &=& 9.6 \times 10^{15} {\rm ~~~[molecule~mly^{-1}\,cm^{-2}]}.
\end{eqnarray*}

\end{appendix}
\end{document}